\documentclass[a4paper]{article}
\usepackage{amsmath}
\usepackage{graphicx}
\usepackage{geometry}
\usepackage{floatrow}
\usepackage{float}
\usepackage{layout}
\usepackage{amssymb} 
\usepackage{multirow}
\usepackage{caption}
\usepackage{subcaption} 
\usepackage{authblk}
\usepackage{hyperref}
\usepackage{indentfirst}
\usepackage{physics}
\usepackage[toc,page]{appendix}
\usepackage{tikz}
\usepackage{cite}
\usetikzlibrary{calc}
\usepackage{relsize}
\tikzset{fontscale/.style = {font=\relsize{#1}}}
\usetikzlibrary{arrows}
\usetikzlibrary{shapes,arrows,shapes.multipart}
\usepackage{circuitikz}

\begin{document}

\newcommand{\beq}{\begin{equation}}
\newcommand{\eeq}{\end{equation}}
\newcommand{\bea}{\begin{eqnarray}}
\newcommand{\eea}{\end{eqnarray}}

\title{\textbf{\huge{Holographic Cosmology on Closed Slices in 2+1 Dimensions}}}
\author{\large{Goncalo Araujo-Regado}\footnote{ga365@cam.ac.uk}}
\affil{DAMTP, University of Cambridge} 

\maketitle

\begin{abstract}
    We apply the framework of Cauchy Slice Holography to the quantization of gravity on closed slices with $\Lambda>0$ (with a focus on $2+1$ dimensions for concreteness). We obtain solutions to the Wheeler-DeWitt equation in a basis of CPT-dual branches. Each branch is a $T^2$-deformed CFT partition function with imaginary central charge. We compute explicit solutions in $2+1$ dimensions in a minisuperspace toy model of pure gravity. This analysis gives us evidence to conjecture a connection between the choice of superposition of branches and the choice of class of geometries to sum over in the gravitational path integral. We further show that, in full quantum gravity on closed slices, bulk unitarity holds, despite the Euclidean holographic field theory not being reflection-positive. We conjecture about the (non)uniqueness of the quantum state of the Universe, in light of this analysis.
\end{abstract}

\tableofcontents

\section{Introduction}

The field of holography has had a tremendous success in the context of the AdS/CFT correspondence. The latter can be interpreted as giving a definition of quantum gravity in asymptotically-AdS spacetimes via a Lorentzian CFT. In recent years a lot of progress has been made in understanding how to do quantum gravity when the boundary has finite intrinsic spatial volume. The boundary description for this seems to involve a deformed field theory on the timelike boundary. The deformation operator away from the CFT is quadratic in the stress tensor (thus irrelevant) and is called the $T^2$-operator in the literature \cite{McGough:2016lol, Hartman:2018tkw}. This approach to quantum gravity in a finite region led to attempts to describe the dS static patch in $2+1$ dimensions holographically via such a deformed theory \cite{Coleman:2021nor}, by combining two such flows, starting from the CFT corresponding to $2+1$-dimensional AdS. The final theory was interpreted to live on a timelike cylinder surrounding an observer worldline inside the patch.

However, by considering holographic descriptions on the timelike boundary, we have not been tackling the problem of bulk quantization directly. This had been the focus of older literature, which tried to solve the Hamiltonian and momentum constraint equations following from the diffeomorphism invariance of the theory \cite{DeWitt:1967yk}. In the context of cosmology, a solution to these equations is a physical state of quantum gravity living on a closed slice $\Sigma$. Solving the Hamiltonian constraint is hard and considerable progress was made in the minisuperspace toy model or extensions thereof with countable number of degrees of freedom \cite{Hartle:1983ai, Halliwell:1984eu}. A lot of discussion has followed in the cosmology literature about what the physically relevant solution to this constraint is \cite{Vilenkin:1984wp, Vilenkin:1986cy, Vilenkin:2018dch, Hartle:2008ng, Feldbrugge:2017mbc, Halliwell:1989dy}. 

But the field of canonical quantization remained largely separate from holographic approaches. It was not long after the discovery of AdS/CFT that conjectures of dS/CFT emerged \cite{Strominger:2001pn}, in which quantum gravity in asymptotically-dS spacetimes was to be defined in terms of a Euclidean CFT living at the future/past asymptotic boundary. A lot of intuition from AdS/CFT has been borrowed since to try to come up with explicit realizations of this conjecture. But a simple process of Wick rotation from a Lorentzian to a Euclidean CFT cannot be the answer, because we do not expect the CFT relevant for dS holography to be reflection-positive, while the CFT for AdS holography is unitary. We will discuss this point more later. In fact, the proposals put forward so far tend to have quite exotic properties \cite{Anninos:2011ui, Hikida:2022ltr}.

The work of \cite{Araujo-Regado:2022gvw} precisely makes the bridge between canonical quantization and holography. In the context of AdS/CFT, it gives quantum gravity states on a bulk slice $\Sigma$ defined in terms of a $T^2$-deformed field theory living on $\Sigma$ and with boundary conditions provided by a field theory state living on $\partial\Sigma$. This framework has the advantage of having a general regime of applicability. In particular, it can be used to do canonical quantization of cosmology on closed slices, by expressing the quantum states in terms of a $T^2$-deformed field theory, where in this case there is no boundary at which to specify field theory data. The main goal of this paper is to explore this holographic description and its conceptual consequences for quantum gravity with $\Lambda>0$ on closed slices. 

Here is a summary of the ideas discussed in this paper. A quantum gravity state living on an abstract slice $\Sigma$ must satisfy the Hamiltonian constraint. All solutions to this equation are spanned by two field theory branches:
\bea
\Psi[g]=A_+Z_+[g]+A_-Z_-[g]=:Z[g],\label{superposition}
\eea
where each of $(Z_\pm)$ is a $T^2$-deformed partition function on $\Sigma$. The starting CFTs for each of them have opposite Weyl anomalies. For $\Lambda>0$ and $2+1$-dimensions this leads to the fact that the CFTs have opposite imaginary central charges. The partition functions start by being CPT duals of each other before any deformation is turned on. This property is preserved along the deformation flow. If we compute the expectation value of the conjugate momentum to $g_{ab}$ (a.k.a. the extrinsic curvature) we get oppositely signed contributions from each branch. This hints at the fact that the branch $Z_+$ encodes future-oriented information, while the branch $Z_-$ encodes past-oriented information, relative to the abstract slice $\Sigma$. From the perspective of the total partition function $Z[g]$ on $\Sigma$ this is related to the phenomenon of spontaneous CPT-symmetry breaking. The field theory living on $\Sigma$ breaks reflection-positivity. 

However, there still is a bulk positive-definite inner product, given by a gravitational path integral. This inner product can be expressed holographically in terms of a field theory partition function. Thus, we have bulk unitarity, despite the lack of reflection-positivity of the dual field theory. The fact that bulk CPT invariance does not seem to be essential for bulk unitarity suggests the absence of a CPT-theorem in quantum gravity. In addition, an important aspect of the holographic encoding of the gravitational path integral is conjectured to be the correspondence between the choice of branch superposition in \eqref{superposition} and the choice of class of geometries in the gravitational path integral. In minisuperspace, we explicitly show how different branch superpositions correspond to different lapse contours.

\begin{figure}
    \centering

\tikzset{every picture/.style={line width=0.75pt}} %set default line width to 0.75pt        

\begin{tikzpicture}[x=0.75pt,y=0.75pt,yscale=-0.75,xscale=0.75]
%uncomment if require: \path (0,466); %set diagram left start at 0, and has height of 466

%Shape: Ellipse [id:dp9805877256817206] 
\draw  [fill={rgb, 255:red, 255; green, 255; blue, 255 }  ,fill opacity=1 ] (220,119.5) .. controls (220,100.45) and (263.2,85) .. (316.5,85) .. controls (369.8,85) and (413,100.45) .. (413,119.5) .. controls (413,138.55) and (369.8,154) .. (316.5,154) .. controls (263.2,154) and (220,138.55) .. (220,119.5) -- cycle ;
%Shape: Polygon Curved [id:ds7273419067814135] 
\draw  [fill={rgb, 255:red, 241; green, 237; blue, 237 }  ,fill opacity=1 ] (220,119.5) .. controls (218.79,111.74) and (226,211) .. (286,230) .. controls (331,246) and (312,328) .. (347,325) .. controls (384,321) and (407,314) .. (379,278) .. controls (355,243) and (406,178) .. (413,119.5) .. controls (416,136) and (367,159) .. (289.79,152.75) .. controls (257.79,147.75) and (241.04,141.75) .. (231.94,135.84) .. controls (222.85,129.94) and (221.4,124.12) .. (220,119.5) -- cycle ;
%Curve Lines [id:da48338574928937894] 
\draw    (430,86) .. controls (404.39,86) and (438.93,106.37) .. (403.66,103.16) ;
\draw [shift={(402,103)}, rotate = 6.01] [color={rgb, 255:red, 0; green, 0; blue, 0 }  ][line width=0.75]    (10.93,-3.29) .. controls (6.95,-1.4) and (3.31,-0.3) .. (0,0) .. controls (3.31,0.3) and (6.95,1.4) .. (10.93,3.29)   ;

% Text Node
\draw (436,78.4) node [anchor=north west][inner sep=0.75pt]    {$\Sigma $};
% Text Node
\draw (202,273) node [anchor=north west][inner sep=0.75pt]   [align=left] {no-boundary\\proposal};
% Text Node
\draw (114,97.4) node [anchor=north west][inner sep=0.75pt]    {$\Psi [ g] =Z[ g]$};

\end{tikzpicture}
    \caption{Pictorial representation of the no-boundary proposal for quantum state of the Universe. The state lives on an abstract slice $\Sigma$ and is obtained by a gravitational path integral over some class of geometries defined on manifolds $\mathcal{M}$ whose only boundary is $\partial\mathcal{M}=\Sigma$. The state obtained corresponds to some linear combination of the form \eqref{superposition}. A priori, there is no preferred time orientation between the no-boundary “boundary" and $\Sigma$. That is to be determined by the linear combination at hand.}
    \label{NB}
\end{figure}
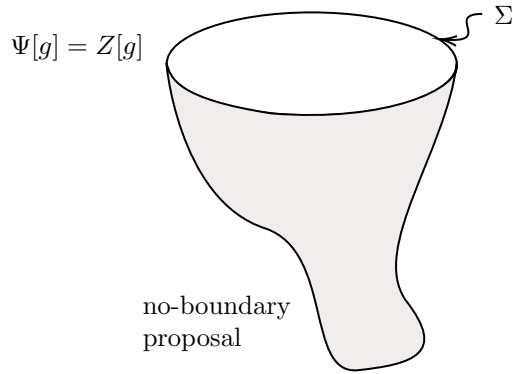

In this paper we also explicitly compute the branches $(Z_\pm)$ in a minisuperspace toy model of pure gravity by integrating the flow. We recover known solutions from the cosmology literature by picking appropriate linear combinations. The bridge between the holographic principle and Wheeler-DeWitt quantization, encapsulated by \eqref{superposition}, is pictorially represented in Figure \ref{NB} for the setting of the no-boundary proposal for the state $\Psi[g]$. The usual dS/CFT statement is recovered in the limit of large volume on $\Sigma$ and where the state \eqref{superposition} is taken to contain only the future/past branch.

The paper is organized as follows. In Section \ref{prescription} we summarize the general framework of Cauchy Slice Holography and derive the deformation operator for pure gravity in $2+1$-dimensions. In Section \ref{mini} we explicitly integrate the deformation flow in a minisuperspace ansatz to obtain the space of solutions to the minisuperspace Hamiltonian constraint. Then, in Section \ref{contour}, we conjecture a correspondence between the choice of contour in the gravitational path integral and the choice of superposition of field theory partition functions and we give some evidence pointing in this direction, using the minisuperspace solutions as a guide. In Section \ref{CPT} we discuss the relationship between bulk CPT invariance, bulk unitarity and (non)reflection-positivity of the field theory. We further comment on how to interpret the apparent uniqueness of the cosmological state, which seems to follow from the formalism. We end with a discussion of these and additional points in Section \ref{discussion}.

\section{Summary of Prescription}\label{prescription}

For now we will be working in general $d+1$ dimensions. We will specialize to $d=2$ momentarily. The Hamiltonian constraint takes the form:
\bea
\mathcal{H}(x)\Psi[g]:=\Bigg\{\frac{16\pi G_N}{\sqrt{g}}:\!\Big(\Pi_{ab}\Pi^{ab}-\frac{1}{d-1}\Pi^2\Big)\!:-\frac{\sqrt{g}}{16\pi G_N}(R-2\Lambda)\Bigg\}\Psi[g]=0,\label{ham}
\eea
and it must hold at every point on $\Sigma$. We are interested in working on closed slices, since we have an eye towards quantization of gravity on global slices of asymptotically dS spacetimes. We will also take $\Lambda>0$. The $::$ symbol denotes normal-ordering and $\Pi^{ab}=-i\frac{\delta}{\delta g_{ab}}$ is the conjugate momentum variable.

We will now follow the construction of Cauchy Slice Holography, introduced at length in \cite{Araujo-Regado:2022gvw}, to solve this constraint. 

As explained in \cite{Freidel:2008sh}, in the region of superspace corresponding to large intrinsic volume of $\Sigma$, a general solution to \eqref{ham} takes the form:
\bea
\lim_{\text{Vol}[g]\to\infty}\Psi[g]= A_{+} e^{+\text{CT}[g] } Z_{+}^{(\infty)}[g] + A_{-} e^{-\text{CT}[g] } Z_{-}^{(\infty)}[g],\label{asymptotics}
\eea
where $\text{Vol}[g]\to\infty$ means with respect to other bulk scales in the theory, like $G_N^{\frac{d}{d-1}}$. Here, $\text{CT}[g]$ is a universal local counterterm. It takes the same form for all states and it only contains terms of scaling dimension $>d$ (which are therefore divergent in this limit). $A_\pm$ are just complex numbers. The other two factors satisfy local anomaly equations:
\bea
\mathcal{W}(x)Z_{\pm}^{(\infty)}[g]=\mp i\mathcal{A}(x)Z_{\pm}^{(\infty)}[g],\label{anomaly}
\eea
where $\mathcal{W}(x)$ is the local Weyl rescaling generator: $\mathcal{W}(x)=2\Pi(x)$; and $\mathcal{A}(x)$ is some local function of $g_{ab}$, with scaling dimension $=d$. It is worth stressing that $\mathcal{A}(x)$ is entirely determined by $\mathcal{H}(x)$, as we will see. They can thus be identified with CFT partition functions, with \emph{opposite} anomalies:
\bea
Z_{\pm}^{(\infty)}[g]=Z_{\pm}^{(\text{CFT})}[g].
\eea
A general solution to the Hamiltonian constraint is thus spanned by two branches.

If we were to apply this to AdS/CFT, one of the branches would be exponentially suppressed relative to the other. So, in the large volume limit only one of them would contribute. That is in fact the premise of AdS/CFT. But if we were to probe regions of finite volume, it is likely that the two branches would become important even in that case. For the purposes of doing a dS/CFT version of holography, the counterterms are oscillatory and so no branch dominates over the other. Both have to be taken into account from the start, even in the limit of the asymptotic boundary. This will in fact turn out to be absolutely crucial to recover a physically reasonable behaviour, as we will see.

Away from this region of superspace, a general state takes the form of linear combination of partition functions deformed away from the asymptotic CFT behaviour. Each branch will have its corresponding deformation:
\bea
Z_\pm [g]=e^{\pm\text{CT}[g]}\left( \text{P} \exp \int_{\epsilon}^\mu\frac{d\lambda}{\lambda}O_\pm(\lambda)\right)Z_{\pm}^{(\text{CFT})}[g],\label{deformation}
\eea
where $O_\pm(\lambda)$ is a local deforming operator composed of terms with scaling dimension $\leq d$. It is updated at each point along the flow parameterized by $\lambda$ (which is why the exponential is path ordered). The flow starts at the UV cutoff $\lambda=\epsilon$ of the CFT and ends at $\lambda=\mu$. This $\mu$ is a new (and the only independent) scale in the problem.  The following convention for identification with bulk scales holds throughout the paper:
\bea
|\mu^{1/d}|=L_\text{(A)dS},
\eea
where 
\bea
\Lambda = (-)\frac{d(d-1)}{2}\frac{1}{L_\text{(A)dS}^2}.
\eea

\begin{figure}[H]
    \centering
\tikzset{every picture/.style={line width=0.75pt}} %set default line width to 0.75pt        

\begin{tikzpicture}[x=0.75pt,y=0.75pt,yscale=-0.8,xscale=0.8]
%uncomment if require: \path (0,437); %set diagram left start at 0, and has height of 437

%Shape: Axis 2D [id:dp3185706067645847] 
\draw  (153,222) -- (510,222)(330,46) -- (330,398) (503,217) -- (510,222) -- (503,227) (325,53) -- (330,46) -- (335,53)  ;
%Straight Lines [id:da031011260600004364] 
\draw [color={rgb, 255:red, 126; green, 211; blue, 33 }  ,draw opacity=1 ]   (329.49,222) -- (491,222) ;
\draw [shift={(417.25,222)}, rotate = 180] [fill={rgb, 255:red, 126; green, 211; blue, 33 }  ,fill opacity=1 ][line width=0.08]  [draw opacity=0] (12,-3) -- (0,0) -- (12,3) -- cycle    ;
%Straight Lines [id:da7615330966088976] 
\draw [color={rgb, 255:red, 245; green, 166; blue, 35 }  ,draw opacity=1 ]   (162,222) -- (329.49,222) ;
\draw [shift={(236.75,222)}, rotate = 0] [fill={rgb, 255:red, 245; green, 166; blue, 35 }  ,fill opacity=1 ][line width=0.08]  [draw opacity=0] (12,-3) -- (0,0) -- (12,3) -- cycle    ;
%Straight Lines [id:da16195770394873343] 
\draw [color={rgb, 255:red, 74; green, 144; blue, 226 }  ,draw opacity=1 ]   (330,222) -- (330,388) ;
\draw [shift={(330,312)}, rotate = 270] [fill={rgb, 255:red, 74; green, 144; blue, 226 }  ,fill opacity=1 ][line width=0.08]  [draw opacity=0] (12,-3) -- (0,0) -- (12,3) -- cycle    ;
%Straight Lines [id:da2915262339939333] 
\draw [color={rgb, 255:red, 144; green, 19; blue, 254 }  ,draw opacity=1 ]   (330,222) -- (330,65) ;
\draw [shift={(330,136.5)}, rotate = 90] [fill={rgb, 255:red, 144; green, 19; blue, 254 }  ,fill opacity=1 ][line width=0.08]  [draw opacity=0] (12,-3) -- (0,0) -- (12,3) -- cycle    ;

% Text Node
\draw (312,17.4) node [anchor=north west][inner sep=0.75pt]    {$\text{Im}( \lambda )$};
% Text Node
\draw (528,212.4) node [anchor=north west][inner sep=0.75pt]    {$\text{Re}( \lambda )$};
% Text Node
\draw (190,186.4) node [anchor=north west][inner sep=0.75pt]    {$\textcolor[rgb]{0.96,0.65,0.14}{d=2,\ \Lambda  >0}$};
% Text Node
\draw (218,310.4) node [anchor=north west][inner sep=0.75pt]    {$\textcolor[rgb]{0.29,0.56,0.89}{d=3,\ \Lambda  >0}$};
% Text Node
\draw (375,185.4) node [anchor=north west][inner sep=0.75pt]    {$\textcolor[rgb]{0.49,0.83,0.13}{\text{all} \ d,\ \Lambda < 0}$};
% Text Node
\draw (373,242.4) node [anchor=north west][inner sep=0.75pt]    {$\textcolor[rgb]{0.49,0.83,0.13}{d=4,\ \Lambda  >0}$};
% Text Node
\draw (223,117.4) node [anchor=north west][inner sep=0.75pt]    {$\textcolor[rgb]{0.56,0.07,1}{d=5,\ \Lambda  >0}$};

\end{tikzpicture}
    \caption{Several deformation contours in the complex-$\lambda$ plane for different dimensions and signs of the cosmological constant. For $\Lambda>0$ the contour rotates counterclockwise by $\pi/2$ with each increment in dimension.}
    \label{deform_contour}
\end{figure}
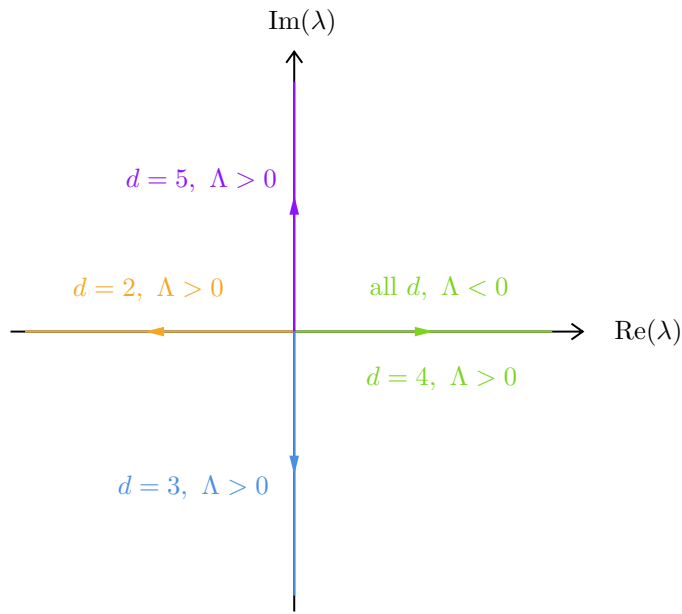

In the complex $\lambda$-plane, the correct deformation contour depends on the sign of $\Lambda$ and on the dimension $d$. In our conventions we have the following pattern shown in Figure \ref{deform_contour}: 

\bea
    \Lambda<0 &\longrightarrow \lambda^{1/d}\in\mathbb{R}_+,\\
    \Lambda>0 &\longrightarrow \lambda^{1/d}\in i\mathbb{R}_+.\label{lambdacontour}
\eea

Of course, the starting CFTs will be wildly different for each $d$. Once we turn on any amount of deformation, the physics will depend on the ratio of the local geometrical scale (determined by $g_{ab}$) and the dimensionful coupling $|\mu^{1/d}|$. In the bulk, the ratio between the cosmological and Planck scales corresponds to some dimensionless parameter in the field theory.

If we define the partition function without the counterterm as:
\bea
Z_\pm[g]=:e^{\pm\text{CT}[g]}Z_\pm '[g],
\eea
then this solves the flow equation:
\bea
\left\{\int_\Sigma\mathcal{W}-id\; O_\pm(\mu)\right\}Z_\pm'[g]=0,\label{floweq}
\eea
by construction.

Let us construct the deforming operator for the case of $d=2$ pure Einstein gravity. We 
first do that for one of the branches, call it $(+)$, and then the other branch is obtained analogously.

We start by dividing the terms in the Hamiltonian operator into relevant, marginal and irrelevant, from the perspective of the $d$-dimensional field theory:
\bea
\mathcal{H}&=&\mathcal{H}^\text{rel}+\mathcal{H}^\text{marg}+\mathcal{H}^\text{irrel}\\
\mathcal{H}^\text{rel}&=&\frac{\sqrt{g}}{16\pi G_N}2\Lambda\\
\mathcal{H}^\text{marg}&=&-\frac{\sqrt{g}}{16\pi G_N}R\\
\mathcal{H}^\text{irrel}&=&\frac{16\pi G_N}{\sqrt{g}}:\!\Big(\Pi_{ab}\Pi^{ab}-\Pi^2\Big)\!:
\eea

We perform a canonical transformation on the phase-space $(g_{ab},\Pi^{ab})$ such that the marginal term gets modified to:
\bea
i\mu^{1/2}\mathcal{H}^\text{marg}\rightarrow -\mathcal{W}-i\frac{\mu^{1/2}}{16\pi G_N}\sqrt{g}R.
\eea
For that we need a conjugate transformation of the form:
\bea
\mathcal{H}\rightarrow e^{-\text{CT}}\;\mathcal{H}\;e^{+\text{CT}},
\eea
with a $d$-dimensional cosmological constant counterterm:
\bea
\text{CT}[g] = +\frac{2}{\mu^{1/2}}\frac{1}{16\pi G_N} \int_{\Sigma} \sqrt{g}.
\eea
Under this, the irrelevant term stays the same, but the relevant term gets modified to:
\bea
\mathcal{H}^\text{rel}\rightarrow\frac{\sqrt{g}}{16\pi G_N}2\left(\Lambda+\frac{1}{\mu}\right).
\eea
In our conventions, in which all the counterterms are to the left of the deformation operator in \eqref{deformation}, this term must be identically zero. Otherwise we will get relevant terms in the anomaly equation \eqref{anomaly} for $Z_\pm^{(\infty)}$, thus violating the premise that $Z_\pm$ satisfied the Hamiltonian constraint. This is equivalent to the convention:
\bea
\Lambda=-\frac{1}{\mu}.
\eea
Picking another identification between the bulk and field theory scales is equivalent to allowing for extra counterterms in \eqref{deformation} \emph{before} deforming, such that we can get rid of any leftover relevant terms in the anomaly equation. This is simply a choice of scheme for the deformation flow in theory space and thus has no physical consequence. We believe our convention to be the cleanest. It also is the same convention used in holographic renormalization.

So after this canonical transformation we have that:
\bea
i\mu^{1/2}\mathcal{H}&\rightarrow& -\mathcal{W}-i\frac{\mu^{1/2}}{16\pi G_N}\sqrt{g}R+i\mu^{1/2}\frac{16\pi G_N}{\sqrt{g}}:\!\Big(\Pi_{ab}\Pi^{ab}-\Pi^2\Big)\!:,\\
&\rightarrow&-\mathcal{W}+2iX_+^{(\mu)},
\eea
where the last line is a definition of the local operator $X_+^{(\mu)}(x)$. This operator itself splits into irrelevant and marginal terms:
\bea
X_+=X_+^\text{marg}+X_+^\text{irrel},
\eea
classified as before. The deformation operator is then the integrated version over the slice:
\bea
O_+(\mu):=\int_\Sigma X_+^{(\mu)}(x).
\eea
Given this deformation operator in \eqref{deformation}, then the anomaly constraint equation is guaranteed to include only the marginal terms:
\bea
\mathcal{W}(x)Z_+^{(\text{CFT})}[g]=- i\left(\frac{\mu^{1/2}}{16\pi G_N}\sqrt{g}R\right)\;Z_+^{(\text{CFT})}[g].
\eea
Using the standard definition of the central charge in $d=2$ we get the following holographic identification:
\bea
c=\frac{3\mu^{1/2}}{2G_N}.
\eea
This is the dimensionless parameter in the field theory which controls the ratio of the bulk scales. This conclusion works for both signs of the bulk cosmological constant. Now if we consider the relevant sign $\Lambda>0$ for cosmology and the associated $\lambda$-contour \eqref{contour} we get an imaginary central charge for the starting CFT:
\bea
\mu^{1/2}=iL_\text{dS}\rightarrow c=i\frac{3L_\text{dS}}{2G_N}.\label{charge}
\eea

In terms of field theory quantities only, the deforming operator and the counterterm take the form:
\bea
O_+(\lambda)&=&\frac{1}{2}\int_\Sigma\left\{-\frac{c}{24\pi}\sqrt{g}R+\frac{24\pi}{c}\lambda \frac{1}{\sqrt{g}}:\!\Big(\Pi_{ab}\Pi^{ab}-\Pi^2\Big)\!:\right\},\label{def}\\
\text{CT}[g]&=&+\frac{c}{12\pi\mu}\int_\Sigma\sqrt{g}.\label{ct}
\eea
So far we have constructed $Z_+[g]$. Notice that there are no boundary conditions to be specified in the definition of this partition function since the slice is closed.

Now to obtain the other branch, we simply swap the sign of the counterterm. This leads to the Hamiltonian constraint turning into a similar form with irrelevant and marginal terms, but with the opposite sign for the Weyl operator:
\bea
i\mu^{1/2}\mathcal{H}\rightarrow \mathcal{W}-2iX_-^{(\mu)},
\eea
where we take this as the definition of local operator $X_-^{(\mu)}$. The deformation operator is similarly defined and we get:
\bea
O_-(\mu):=\int_\Sigma X_-^{(\mu)}(x)=-O_+(\mu).
\eea
This defines the branch $Z_-[g]$. The starting CFT corresponding to this branch now solves the opposite anomaly constraint equation:
\bea
\mathcal{W}(x)Z_-^{(\text{CFT})}[g]=+ i\left(\frac{\mu^{1/2}}{16\pi G_N}\sqrt{g}R\right)\;Z_-^{(\text{CFT})}[g].
\eea

It is worth stressing that for $\Lambda>0$ (and hence for $c\in i\mathbb{R}_+$ and $\mu<0$ in our conventions) we get that the two branches are actually complex conjugates of each other:
\bea
Z_-[g]=\left(Z_+[g]\right)^*.
\eea
This is crucial to obtaining a physically sensible bulk picture, as will be explored below.

\pagebreak

\section{Minisuperspace Solutions}\label{mini}

In general, computing the deformation flow explicitly is hard, because it involves integrating a functional differential operator. However, we can make progress in a minisuperspace ansatz for the metric. This is only a toy model for quantum gravity, which ignores most of the degrees of freedom. However, we hope it can give us some hints about general lessons valid in full quantum gravity. We also take the global slice to have the topology of a sphere, $\Sigma=S^2$ (we could consider other topologies). Thus the metric takes the form:
\bea
g_{ab}=a^2\Omega_{ab}=:q\;\Omega_{ab},
\eea
where $a$ is the sphere radius, $\Omega_{ab}$ is the unit radius spherical metric and $q$ is defined above.

We proceed as follows: we start by finding the functional dependence of $Z_\pm^{(\text{CFT})}$ on $q$; we write the deformation operator in minisuperspace and solve the flow equation \eqref{floweq}; we end by adding the counterterm. The end result will be a solution to the minisuperspace Hamiltonian constraint.

An anomalous CFT has a UV cutoff scale $|\epsilon|$. In our conventions, it has mass dimension $[|\epsilon|]=-2$. Because we want to identify it with the starting point of the flow, where $\lambda=\epsilon$, we have that $\epsilon<0$.

The integrated Weyl anomaly equation \eqref{anomaly} tells us how the existence of said UV cutoff induces an RG flow in the CFT. It encodes the fact that:
\bea
\left\{q\frac{\partial}{\partial q}+\epsilon\frac{\partial}{\partial\epsilon}\right\}Z_\pm^{(\text{CFT})}=0,
\eea
which implies that the partition function only depends on the ratio of physical scales:
\bea
Z_\pm^{(\text{CFT})}=Z_\pm^{(\text{CFT})}(q/\epsilon).
\eea
The following relations hold:
\bea
\int_{S^2}\mathcal{W}&=&-2iq\frac{\partial}{\partial q},\\
\pm i\int_{S^2}\mathcal{A}&=&-2i\epsilon\frac{\partial}{\partial\epsilon}.
\eea
As derived above, the anomaly takes the form:
\bea
\mathcal{A}=-\frac{c}{24\pi}\sqrt{g}R.
\eea
The Gauss-Bonnet theorem tells us that:
\beq
\int_{S^2}\sqrt{g}R=4\pi(2-2g)=8\pi.
\eeq
So the integrated anomaly equation becomes:
\bea
\frac{\partial \log{Z_\pm^{(\text{CFT})}(q)}}{\partial \log{q}}=\pm\frac{c}{6},
\eea
and the solution is, up to an overall normalization:
\bea
Z_\pm^{(\text{CFT})}(q/\epsilon)\propto\left(\frac{q}{\epsilon}\right)^{\pm c/6}.
\eea
Because $c\in i\mathbb{R}_+$, the two CFT branches are complex conjugates of each other in their functional dependence on $q$.  
The counterterm $\eqref{ct}$ takes the form:
\bea
\text{CT}(q/\mu) =+\frac{c}{3}\left(\frac{q}{\mu}\right).
\eea

Taking the large volume limit relative to other bulk scales corresponds to taking $q\gg|\mu|$. Equivalently, since the functional dependence is only on the ratio $q/\mu$, we could take the limit of small deformation away from the CFT, i.e. we could take $\mu\to\epsilon$.
The large volume limit of any solution to the minisuperspace Hamiltonian constraint in $d=2$ is thus:
\bea
\lim_{\mu\to\epsilon}\Psi(q/\mu)=A_+\;e^{+\frac{c}{3}\left(\frac{q}{\epsilon}\right)}\left(\frac{q}{\epsilon}\right)^{+ c/6}+A_-\;e^{-\frac{c}{3}\left(\frac{q}{\epsilon}\right)}\left(\frac{q}{\epsilon}\right)^{- c/6}.
\eea

Now we perform the deformations on each branch so that we obtain the quantum state for finite volume of the slice. In the minisuperspace ansatz, and imposing the normal-ordering prescription (which puts all derivatives to the right), we get the following kinetic term:
\bea
:\!\left(\Pi^{ab} \Pi_{ab}\ - \Pi^2\right)\!:&=& \frac{1}{2}\left(\frac{\sqrt{\Omega}}{4\pi}\right)^2q^2\frac{\partial^2}{\partial q^2}.\label{Tsquared}
\eea
Thus, using \eqref{def}:
\bea
O_\pm(\lambda)=\pm \frac{1}{2}\left\{-\frac{c}{3}+\frac{3}{c}\lambda \;q\frac{\partial^2}{\partial q^2}\right\}
\eea
We now get the flow equation:
\bea
\left\{-2iq\frac{\partial}{\partial q}-(\pm)i\left(-\frac{c}{3}+\frac{3}{c}\mu \;q\frac{\partial^2}{\partial q^2}\right)\right\}Z'_\pm(q/\mu)=0.
\eea
Varying $q$ while keeping $\mu$ fixed is equivalent to varying the ratio $q/\mu$. We define the dimensionless variable:
\bea
u:=-\frac{q}{\mu}>0,
\eea
since $\mu<0$ in our conventions. We also write $c=i|c|$. The flow equation becomes:
\bea
\left\{u\frac{d^2}{du^2}\mp i\frac{2|c|}{3}u\frac{d}{du}-\left(\frac{|c|}{3}\right)^2\right\}Z'_\pm(u)=0.
\eea
We notice that the differential operators corresponding to each branch are complex conjugates of each other. We will now find the general solutions to these equations and demand that they have the correct CFT limit when $u\to\infty$. This will specify the solutions up to a constant, which is just a normalization factor. 

The solutions to this equation are:
\bea
Z'_\pm(u)= \alpha_\pm \;(\pm) i\frac{2|c|}{3}u\;_1F_1\left(1\mp i\frac{|c|}{6},2,\pm i\frac{2|c|}{3}u\right)+\beta_\pm\; e^{\pm i\frac{2|c|}{3}u}\;U\left(\pm i\frac{|c|}{6},0,\mp i\frac{2|c|}{3}u\right),
\eea
where $_1F_1$ and $U$ are standard hypergeometric functions. We pick $(\alpha_\pm,\beta_\pm)$ such that they asymptote to:
\bea
\lim_{u\to\infty}Z'_\pm(u)\propto u^{\pm i\frac{|c|}{6}}+\text{ subleading.}
\eea
We need:
\bea
\beta_\pm=\alpha_\pm\;\frac{-e^{-\frac{\pi|c|}{6}}}{\Gamma\left(1\mp i\frac{|c|}{6}\right)}.
\eea
Picking a convenient normalization, the full solution for each branch becomes, including counterterms:
\bea\label{solution}
Z_\pm(u)=e^{\mp i\frac{|c|}{3}u}\;\Bigg\{&+&\frac{2|c|}{3}u\;_1F_1\left(1\mp i\frac{|c|}{6},2,\pm i\frac{2|c|}{3}u\right)\\
&\pm&i\frac{e^{-\frac{\pi|c|}{6}}}{\Gamma\left(1\mp i\frac{|c|}{6}\right)}\; e^{\pm i\frac{2|c|}{3}u}\;U\left(\pm i\frac{|c|}{6},0,\mp i\frac{2|c|}{3}u\right)\Bigg\}.
\eea
They are indeed complex conjugates of each other.

\begin{figure}[H]
    \centering
    \includegraphics{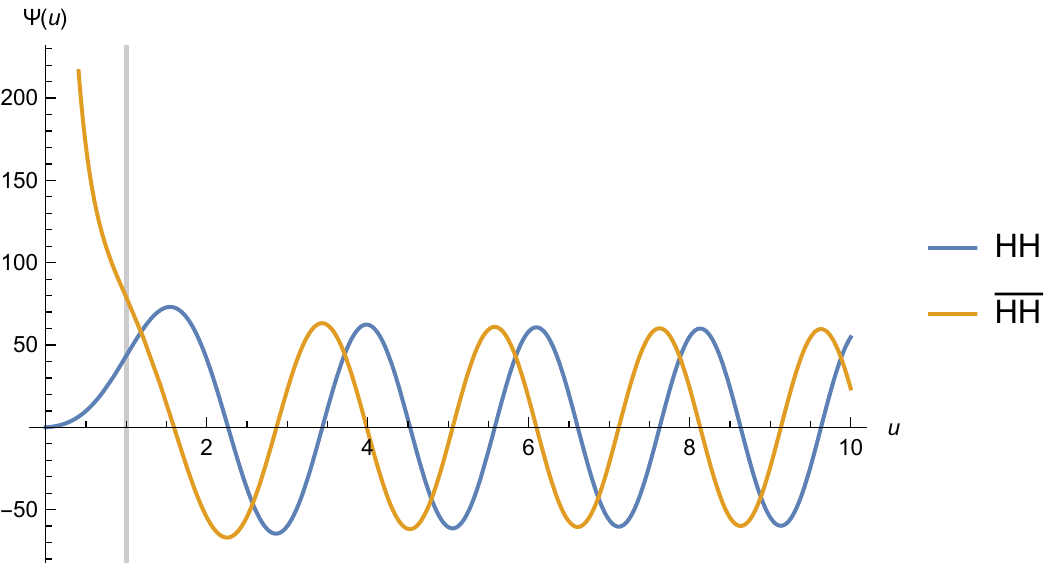}
    \caption{The two solutions \eqref{HH} and \eqref{barHH} for a value of $|c|=10$. The vertical line represents the transition point at $u=1$ beyond which the oscillatory behaviour begins. The two solutions are $\pi/2$ out of phase.}
    \label{both}
\end{figure}

The space of solutions to the minisuperspace Hamiltonian constraint is then:
\bea
\text{span}_\mathbb{C}\;\{Z_\pm\}.
\eea
Let us take three instructive solutions:
\bea
\Psi_\text{HH}(u):&=&\frac{Z_+(u)+Z_-(u)}{2},\label{HH}\\
\Psi_\text{Vilenkin}(u):&=&i Z_-(u),\label{Vilenkin}\\
\Psi_{\overline{\text{HH}}}(u):&=&\frac{Z_+(u)-Z_-(u)}{2i},\label{barHH}
\eea
where the labels “HH", “Vilenkin" and “$\overline{\text{HH}}$" are merely suggestive of their familiar functional dependence. Figure \ref{both} shows a plot of \eqref{HH} and \eqref{barHH} and we can see that the first one shows the expected behaviour of the minisuperspace solution obtained by Hartle and Hawking (HH) \cite{Hartle:1983ai}, with the exponential suppression towards small $u$. The other agrees with the real part of the Vilenkin solution \cite{Vilenkin:1984wp}. We call it “$\overline{\text{HH}}$" because it is obtained via a dual set of saddle-points in the path integral formulation, as we will explain later. The actual full Vilenkin solution \eqref{Vilenkin} posits a tunnelling process from “nothing". It has the exponential enhancement shown in Figure \ref{both} for small $u$ but it includes only an “outgoing" mode for large $u$, thus explicitly breaking time-reversal symmetry (which is directly related to it being a  complex solution). We will discuss these points further later. 

There are two equivalent ways of obtaining the HH and Vilenkin solutions from the gravitational side. One is by solving directly (via the WKB method) the minisuperspace Hamiltonian constraint, which can be reinterpreted as the time-independent Schrodinger equation for a particle of zero energy in the background of a potential barrier. The HH solution then represents the wavefunction of a particle bouncing off the barrier located in the region $0<u<1$. There is an exponential suppression in the classically forbidden region. Outside the barrier we have a standing wave. The Vilenkin solution has a non-zero amplitude in the classically forbidden region and only has the outgoing mode outside. It is interpreted as the wavefunction for the particle to tunnel into the classically allowed region from the classically forbidden region. The point $u=0$ corresponds to a Universe of zero spatial volume. Thus this wavefunction encodes the seeming appearance of an expanding Universe from “nothing". 

The other equivalent way is by defining each solution in terms of a gravitational path integral with a given lapse contour (much more discussion on the contour issue to follow in Section \ref{contour}). Both of them correspond to a “no-boundary" proposal in which one sums over geometries with the only boundary condition specified by the argument of the wavefunction. The Vilenkin solution is obtained by summing over Lorentzian geometries which “pinch" to zero size in the past. So the spatial slice on which we are evaluating the wavefunction is always to the future of the “nothing" slice. In this approach, time-reversal is explicitly broken by the choice of contour. There is still much debate about the contour corresponding to the HH solution. We will argue in Section \ref{contour} against the contour specified in the HH paper itself. There they defined the path integral to be over Euclidean geometries which are regular at the “no-boundary". However, this has the well-known issues of convergence due to the conformal mode problem \cite{Gibbons:1978ac}. To us it seems much more natural that the contour be chosen along Lorentzian geometries, now with both time orientations of the spatial slice relative to the “no-boundary". The HH state is thus time-reversal symmetric by construction, also manifested in its reality (this connection will be explored in Section \ref{CPT}). For both cases, the actual wavefunction is computed in the saddle-point approximation. Thus, the functional dependence of $\Psi$ boils down to which saddles we can deform each contour to pass through. As the argument of $\Psi$ varies there is a transition point (at $u=1$) where the saddle geometries change from Euclidean to complex. The oscillatory behaviour in Figure \ref{both} for $u>1$ corresponds precisely to these complex saddles.

\begin{figure}[H]
    \centering
    \includegraphics{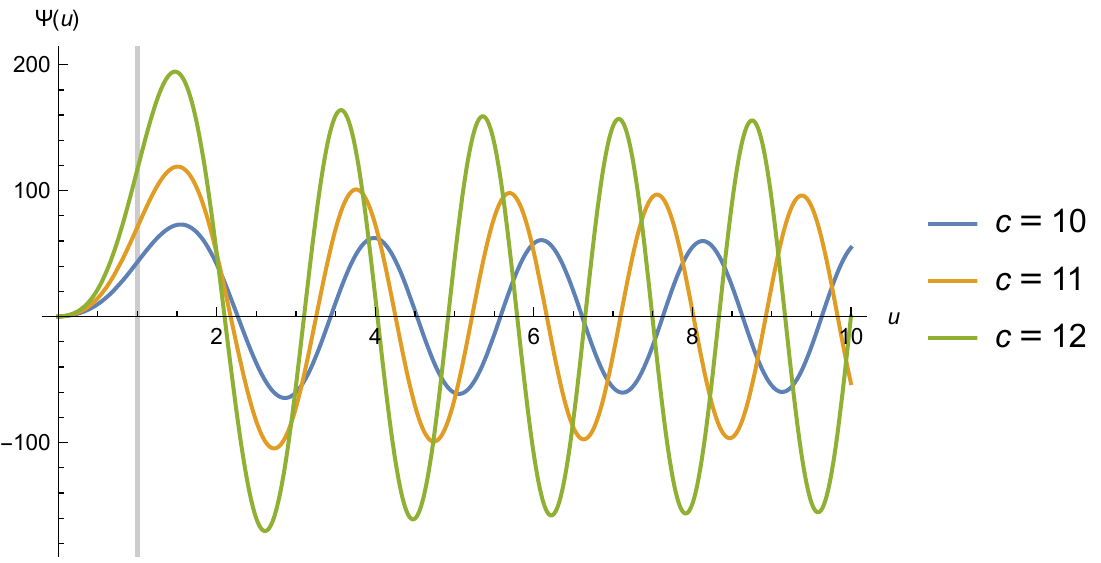}
    \caption{Evolution of the “HH"-like solution with $|c|$.}
    \label{HH_fig}
\end{figure}

\begin{figure}[H]
    \centering
    \includegraphics{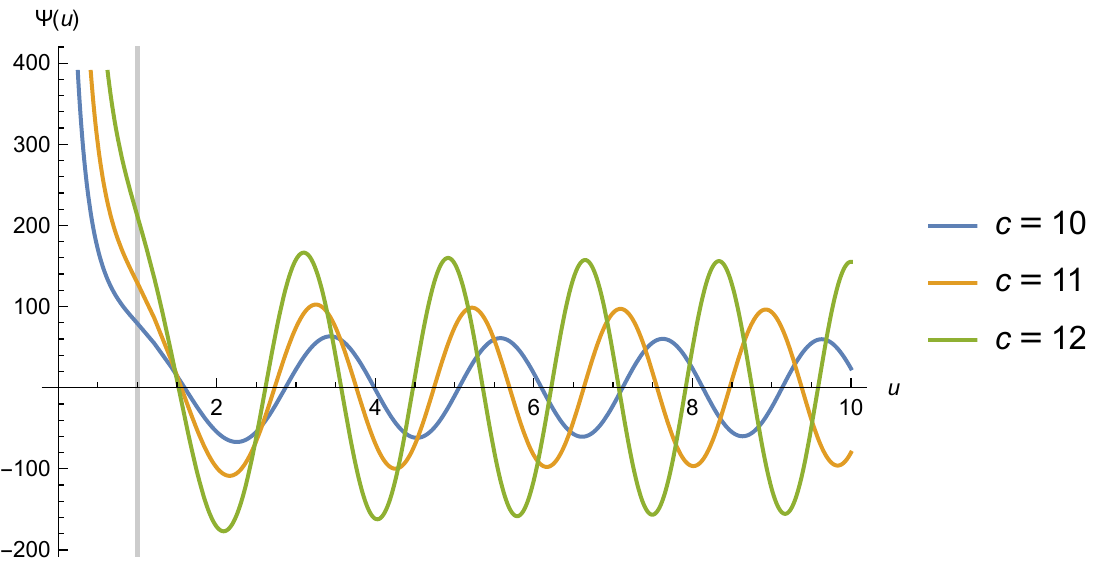}
    \caption{Evolution of the “$\overline{\text{HH}}$"-like solution with $|c|$.}
    \label{Vilenkin_fig}
\end{figure}

We might also wonder how the solutions evolve as we vary the free parameter $|c|$. This is shown in Figures \ref{HH_fig} and \ref{Vilenkin_fig}. We see that the period of the oscillations decreases while the amplitude increases as we increase $|c|$. This is in line with the expectation that increasing the parameter $|c|$ of the field theory corresponds to approaching the semi-classical behaviour of quantum gravity, as suggested by \eqref{charge}. Note that the transition point is always the same.

Working in minisuperspace turned the functional flow equation \eqref{floweq} into an ODE that could be solved explicitly. If we were to turn on all the gravitational degrees of freedom, we would have needed to define the operator in \eqref{Tsquared} via point-splitting. We do not have this problem in minisuperspace since the operator is just a partial derivative. Thus, the operator is well-defined at all scales and, in particular, we can trust it all the way into the UV, i.e. when $q\to 0$, even though it is an irrelevant operator in the field theory. This is translated in the fact that we can trust the solution to the ODE for all values of $q$. This might give us some hints about the UV completion of the $T^2$-deformed field theory. We leave this analysis for future work.

\section{A Conjecture for the Path Integral Contour}\label{contour}

In the path integral approach to quantum gravity, a quantum state on a codimension-1 manifold $\Sigma$ can be defined via:
\bea
\Psi[g]=\int_{\{\textbf{g}\in\mathcal{C}:\textbf{g}|_\Sigma=g\}/\sim} \mathcal{D}[\textbf{g}]\;e^{+iI[\textbf{g}]},\label{pathintegral}
\eea
where \textbf{g} is the bulk spacetime metric. We sum over all equivalency classes of geometries from a given set $\mathcal{C}$ such that the induced metric on $\Sigma$ is $g$. The equivalence relation $\sim$ is provided by bulk diffeomorphisms. There is an induced measure, $\mathcal{D}[\textbf{g}]$, on the quotient space. For appropriate choices of $\mathcal{C}$ such a quantum state is guaranteed to satisfy the momentum and Hamiltonian constraints on $\Sigma$ (we will discuss such choices momentarily). Crucially, the state $\Psi[g]$ is in one-to-one correspondence with the set $\mathcal{C}$ of histories we sum over. Many proposals for this $\mathcal{C}$ have been provided in the literature and there is ongoing debate as to which one is the most physically relevant. We will not attempt to answer this question here. However, we will provide evidence for how the choice of $\mathcal{C}$ manifests itself on the field theory side.

It is important to note that specifying the set of allowed histories is an integral part of the definition of a physical theory, as are the action and the measure factor. For instance, we define the transition amplitude between two basis elements $\ket{g_1}$ on a slice $\Sigma_1$ and $\ket{g_2}$ on a slice $\Sigma_2$ via the gravitational path integral (we will later refer to this as the \emph{dynamical} inner product):
\bea
\braket{g_1}{g_2}_\text{dyn}:=\sum_\mathcal{M} \int_{\{\textbf{g}\in\mathcal{C}:\textbf{g}|_{\Sigma_1}=g_1,\textbf{g}|_{\Sigma_2}=g_2\}/\sim} \mathcal{D}[\textbf{g}]\;e^{+iI[\textbf{g}]},\label{dynamical}
\eea
where we sum over all manifolds $\{\mathcal{M}\}$ with boundary given by $\partial\mathcal{M}=\Sigma_1\cup\Sigma_2$. The specification of $\mathcal{C}$ is a crucial aspect of the definition.\footnote{In the context of closed cosmologies, suppose that $\mathcal{C}$ restricted all geometries to contain a spatial slice of a given fixed volume (after having picked a gauge). Then the transition amplitude between two Universe sizes larger than this fixed volume can only include geometries that first contract and then expand. This clearly leads to a different result than if we had no restriction at all, since in that case we could have had purely expanding/contracting geometries interpolating between the two.} Different choices of $\mathcal{C}$ lead to different choices of inner product (assuming that there is more than one choice which gives positive-definiteness) and hence different probabilities and physical predictions.

In fact, a specification of $\mathcal{C}$ not only defines a given quantum gravity theory, but also uniquely defines a quantum state within that theory, given some “initial condition" (we use scare quotes because the relative time orientation between the two boundaries is to be determined by $\mathcal{C}$ itself). That is because a state of the form \eqref{pathintegral} can be written as the overlap:
\bea
\Psi[g]=\braket{g}{\text{{b.c.}}}_\text{dyn},\label{overlap}
\eea
where $\ket{\text{b.c.}}$ represents some other boundary condition. For example, in the case of closed cosmology (where $\Sigma$ has no boundary), the no-boundary proposal of \cite{Hartle:1983ai,Vilenkin:1984wp} takes $\ket{\text{b.c.}}=\ket{\emptyset}$, where $\ket{\emptyset}$ stands for a slice of zero volume. Thus it is an amplitude of the form \eqref{dynamical} with whatever choice of $\mathcal{C}$ we had previously decided on. We could also define the state corresponding to an initial condition $\ket{\text{b.c.}}=\ket{g_0}$. Other types of “conjugate" boundary conditions are also possible, like $\ket{\text{b.c.}}=\ket{\Pi=0}$. This can be expanded in the basis $\{\ket{g}\}$ and so is again defined by \eqref{dynamical} and its choice of $\mathcal{C}$. The point is that it does not make sense to freely change our choice of $\mathcal{C}$ for different states. Such a change would correspond to changing the quantum gravity theory we care considering. At most, we can change the initial condition to get different states. 

But if we look at equation \eqref{superposition}, where exactly is the freedom in specifying $\ket{\text{b.c.}}$? And where does the information about $\mathcal{C}$ go? We will argue that the answer to the latter is that it manifests itself in the choice of superposition of the two CPT-dual field theory branches.

\subsection{Open vs Closed Cosmologies}

The holographic principle posits that a quantum field theory is a dual description of quantum gravity. This means that the \emph{same} field theory can be used to describe both the dynamics and \emph{all} states of quantum gravity. Given a quantum field theory, the object:
\bea
Z[g],
\eea
defined on a closed manifold $\Sigma$, is determined. This is a functional of $g$, but it has no information about the different possible entries for $\ket{\text{b.c.}}$ in \eqref{overlap}. Of course, the full quantum gravity state is a superposition of two such field theory objects $(Z_\pm[g])$, so one might wonder if the knowledge of $\ket{\text{b.c.}}$ is in the choice of linear combination.\footnote{Notice how defending such a viewpoint seemingly requires a modification of the holographic principle to saying that \emph{two} quantum field theories give a dual description of quantum gravity. However, because, as we will argue, the two partition functions are actually CPT duals of each other, the physics of one is determined entirely by the physics of the other and so we are not introducing new field theory degrees of freedom in the description. In addition, and perhaps more importantly, the two partition functions can be understood as emerging from a phase transition starting from a \emph{single} partition function of a hypothetical UV-complete field theory. See Section \ref{discussion} for an elaboration of this point.} To understand why this is not the case, we will temporarily consider a case in which the spatial slice $\Sigma$ has a non-trivial boundary. 

\begin{figure}
    \centering
    \begin{subfigure}[b]{0.4\textwidth}
        \centering
        \tikzset{every picture/.style={line width=0.75pt}} %set default line width to 0.75pt        

\begin{tikzpicture}[x=0.75pt,y=0.75pt,yscale=-0.75,xscale=0.75]

%Shape: Arc [id:dp18207947443503403] 
\draw  [draw opacity=0] (424.24,179.67) .. controls (424.24,179.73) and (424.24,179.79) .. (424.24,179.85) .. controls (424.28,194.25) and (382.36,206.02) .. (330.62,206.15) .. controls (279.32,206.27) and (237.62,194.9) .. (236.89,180.67) -- (330.56,180.08) -- cycle ; \draw   (424.24,179.67) .. controls (424.24,179.73) and (424.24,179.79) .. (424.24,179.85) .. controls (424.28,194.25) and (382.36,206.02) .. (330.62,206.15) .. controls (279.32,206.27) and (237.62,194.9) .. (236.89,180.67) ;  
%Shape: Arc [id:dp2658183196997167] 
\draw  [draw opacity=0][dash pattern={on 4.5pt off 4.5pt}] (236.89,180.67) .. controls (236.89,180.55) and (236.88,180.43) .. (236.88,180.31) .. controls (236.84,163.74) and (278.75,150.2) .. (330.5,150.06) .. controls (382.24,149.93) and (424.22,163.25) .. (424.27,179.82) .. controls (424.27,180.06) and (424.26,180.31) .. (424.24,180.55) -- (330.58,180.06) -- cycle ; \draw  [dash pattern={on 4.5pt off 4.5pt}] (236.89,180.67) .. controls (236.89,180.55) and (236.88,180.43) .. (236.88,180.31) .. controls (236.84,163.74) and (278.75,150.2) .. (330.5,150.06) .. controls (382.24,149.93) and (424.22,163.25) .. (424.27,179.82) .. controls (424.27,180.06) and (424.26,180.31) .. (424.24,180.55) ;  
%Shape: Arc [id:dp692536611189474] 
\draw  [draw opacity=0] (239.07,185.82) .. controls (228.5,169.07) and (222.41,149.21) .. (222.48,127.94) .. controls (222.65,68.27) and (271.17,20.04) .. (330.83,20.22) .. controls (390.5,20.4) and (438.73,68.91) .. (438.55,128.58) .. controls (438.49,148.84) and (432.86,167.78) .. (423.1,183.96) -- (330.51,128.26) -- cycle ; \draw   (239.07,185.82) .. controls (228.5,169.07) and (222.41,149.21) .. (222.48,127.94) .. controls (222.65,68.27) and (271.17,20.04) .. (330.83,20.22) .. controls (390.5,20.4) and (438.73,68.91) .. (438.55,128.58) .. controls (438.49,148.84) and (432.86,167.78) .. (423.1,183.96) ;  
%Shape: Arc [id:dp9118574417193872] 
\draw  [draw opacity=0][dash pattern={on 0.84pt off 2.51pt}] (423.68,182.99) .. controls (404.76,215.13) and (369.7,236.58) .. (329.72,236.29) .. controls (290.26,236) and (255.9,214.59) .. (237.28,182.87) -- (330.51,128.26) -- cycle ; \draw  [dash pattern={on 0.84pt off 2.51pt}] (423.68,182.99) .. controls (404.76,215.13) and (369.7,236.58) .. (329.72,236.29) .. controls (290.26,236) and (255.9,214.59) .. (237.28,182.87) ;  
%Curve Lines [id:da8079796317337229] 
\draw    (201.79,204.97) .. controls (241.39,175.27) and (214.62,224.81) .. (252.89,196.7) ;
\draw [shift={(254.07,195.82)}, rotate = 143.13] [color={rgb, 255:red, 0; green, 0; blue, 0 }  ][line width=0.75]    (10.93,-3.29) .. controls (6.95,-1.4) and (3.31,-0.3) .. (0,0) .. controls (3.31,0.3) and (6.95,1.4) .. (10.93,3.29)   ;

% Text Node
\draw (324,117.4) node [anchor=north west][inner sep=0.75pt]    {$\Sigma $};
% Text Node
\draw (163,199.4) node [anchor=north west][inner sep=0.75pt]    {$\psi _{\text{QFT}}$};

\end{tikzpicture}
        \caption{Open spatial slice. A field theory state, $\psi_\text{QFT}$, is specified as a boundary condition for the partition function and is part of its definition.}
        \label{open}
    \end{subfigure}
\hfill
\begin{subfigure}[b]{0.4\textwidth}
        \centering

\tikzset{every picture/.style={line width=0.75pt}} %set default line width to 0.75pt        

\begin{tikzpicture}[x=0.75pt,y=0.75pt,yscale=-0.75,xscale=0.75]

%Shape: Arc [id:dp13505588839808813] 
\draw  [draw opacity=0] (259.07,205.82) .. controls (248.5,189.07) and (242.41,169.21) .. (242.48,147.94) .. controls (242.65,88.27) and (291.17,40.04) .. (350.83,40.22) .. controls (410.5,40.4) and (458.73,88.91) .. (458.55,148.58) .. controls (458.49,168.84) and (452.86,187.78) .. (443.1,203.96) -- (350.51,148.26) -- cycle ; \draw   (259.07,205.82) .. controls (248.5,189.07) and (242.41,169.21) .. (242.48,147.94) .. controls (242.65,88.27) and (291.17,40.04) .. (350.83,40.22) .. controls (410.5,40.4) and (458.73,88.91) .. (458.55,148.58) .. controls (458.49,168.84) and (452.86,187.78) .. (443.1,203.96) ;  
%Shape: Arc [id:dp1453558722172087] 
\draw  [draw opacity=0] (443.68,202.99) .. controls (424.76,235.13) and (389.7,256.58) .. (349.72,256.29) .. controls (310.26,256) and (275.9,234.59) .. (257.28,202.87) -- (350.51,148.26) -- cycle ; \draw   (443.68,202.99) .. controls (424.76,235.13) and (389.7,256.58) .. (349.72,256.29) .. controls (310.26,256) and (275.9,234.59) .. (257.28,202.87) ;  

% Text Node
\draw (339,142.4) node [anchor=north west][inner sep=0.75pt]    {$\Sigma $};

\end{tikzpicture}
        \caption{Closed spatial slice. There is no boundary where to specify a field theory state. The partition function does not depend on any such data.}
        \label{closed}
    \end{subfigure}

    \caption{Contrast between the two scenarios of open and closed spatial slice $\Sigma$.}
    \label{openvsclosed}
\end{figure}
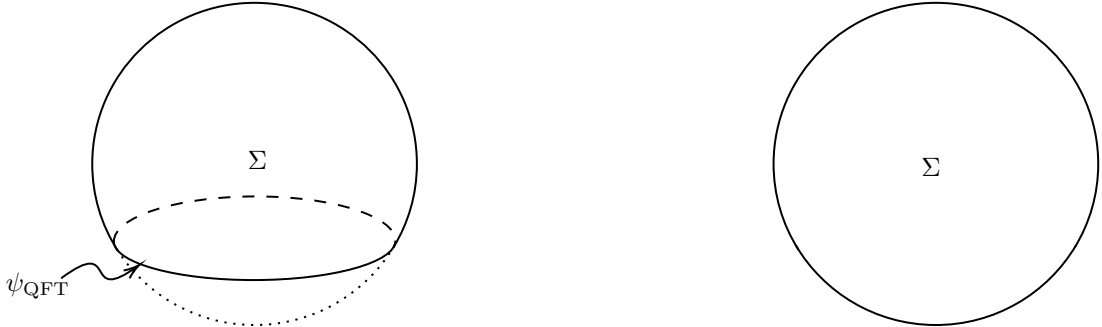

There, the partition function of a field theory living on $\Sigma$ depends not only on the background geometry $g$ but also on some field theory boundary conditions specified at $\partial\Sigma$. There is a field theory Hilbert space, $H_\text{QFT}$, living on $\partial\Sigma$ (since this is a codimension-1 cut of $\Sigma$), and so the specification of said boundary conditions is actually provided by a field theory state $\psi_\text{QFT}\in H_\text{QFT}$, as shown in Figure \ref{openvsclosed}. The correct field theory object to use is thus:
\bea
Z[g,\psi_\text{QFT}].
\eea
The functional dependence of $Z$ on $g$ depends on the choice of $\psi_\text{QFT}$. This is where the information about $\ket{\text{b.c.}}$ is encoded. Different field theory states correspond to different quantum gravity states. This correspondence between states is explored in detail in \cite{Araujo-Regado:2022gvw} in the context of AdS/CFT.

So it seems that in the case of closed spatial slices, there is no place to encode information about $\ket{\text{b.c.}}$. The object $Z[g]$ encodes the information about a single quantum gravity state. But which one? We will attempt to address this question momentarily, but first we will argue that the choice of linear combination of $(Z_\pm)$ is actually related to the choice of set of histories $\mathcal{C}$ and not, as we have just seen, to the choice of $\ket{\text{b.c.}}$.

\subsection{The Lapse Contour}
We saw in Section \ref{mini}, in a minisuperspace toy model, that different branch superpositions in \eqref{superposition} led to different solutions to the same Hamiltonian constraint. We also briefly explained how these well-known solutions can be obtained via a saddle-point approximation in the path integral formulation of the theory and how the choice of lapse contour determines which saddle geometries can contribute. Let us review how this works in minisuperspace.

The spacetime metric takes the form:
\bea
    ds^2=-N(t)^2dt^2+q(t)\;\Omega_{ab}dx^adx^b,
\eea
where $t$ simply parameterizes the history worldline. Here $N(t)$ is the lapse\footnote{A spacetime is a collection $(\mathcal{M},\textbf{g},\xi)$ of a manifold $\mathcal{M}$, with a metric structure \textbf{g} and a time orientation given by some representative timelike vector field $\xi$. In our minisuperspace toy model, we will interpret swapping the sign of the lapse $N\mapsto -N$ as flipping the time orientation of the spacetime. The spacetime metric is invariant under this. When $N\in\mathbb{R}$, the spacetime is Lorentzian and so $N\mapsto -N$ flips future and past. When $N\in i\mathbb{R}$, the spacetime is Euclidean and so $N\mapsto -N$ flips “left" and “right". A complex $N$ corresponds to a complex geometry. We will see shortly that such a geometry can always be deformed into a purely Euclidean geometry being glued to a purely Lorentzian one.} and the function $q$ embeds the worldline into target space:
\bea
q:[0,1]\to \text{Minisuperspace}\cong\mathbb{R}_+
\eea
After imposing this ansatz, the leftover consequence of general covariance of the full gravity theory is reparameterization invariance of the worldline theory:
\bea
t\mapsto \Tilde{t}(t).
\eea

For the case of interest, with no spatial boundary, the Hamiltonian takes the form, in $d=2$:
\bea
p_q\Dot{q}-L=NH=N\int_{S^2}\mathcal{H}=N\left(-2G_Nq\;p_q^2+\frac{1}{2G_N}(\Lambda q-1)\right),
\eea
where $p_q$ is the conjugate momentum to $q$.\footnote{We can see here explicitly that the Hamiltonian constraint equation, $H\Psi(q)=0$, is equivalent to the Schrodinger equation for a non-relativistic particle in a potential of the form $V(q)\sim \frac{1}{q}-\Lambda$. For a particle with energy $E=0$, there is a potential barrier starting at $q=1/\Lambda\iff u=1$, as expected.}

In order to find the measure on the quotient space, we use the Faddeev-Popov prescription. This is explained in various references: \cite{Teitelboim:1981ua}, \cite{PhysRevD.38.2468}. A good gauge-fixing condition is:
\bea
\Dot{N}=0.
\eea
This does not kill physical modes and there is no leftover redundancy in the system. Thus, the quantum state would be given by:
\bea
\Psi(q)=\int \prod_t \left(dq(t)\;dp_q(t)\right)\;\int_{\mathcal{C}}\mathcal{D}N(t)\;e^{+i\int_0^1 dt\left(p_q\Dot{q}-NH\right)}\;\delta(\Dot{N})\;\Delta_\text{FP},\label{state}
\eea
with given boundary conditions for $q$ at the endpoints of $[0,1]$ and where the measure on phase-space is the usual Liouville measure (which is reparameterization invariant) and $\Delta_\text{FP}$ is the Jacobian induced by restricting to the gauge slice. It can be shown that:
\bea
\Delta_\text{FP}=\Delta t =1.
\eea
After imposing the delta function condition:
\bea
N(t)=N,
\eea
%we get:
%\bea
%\Psi(q)&=&\int_{\mathcal{C}}dN\;\int_\text{b.c.}^{q(1)=q} \prod_{t=0}^1 \left(dq(t)\;dp_q(t)\right)\;e^{+i\int_0^1 dt\left(p_q\Dot{q}-NH\right)},\\
%&=&\int_\text{b.c.}^{q(1)=q} \prod_{t=0}^1 \left(dq(t)\;dp_q(t)\right)\;e^{+i\int_0^1 dt\; p_q\Dot{q}}\left(\int_{\mathcal{C}}dN\;e^{-iN\int_0^1 dt\; H}\right).
%\eea
we can define the proper time variable:
\bea
\tau:=Nt,
\eea
such that:
\bea
\Psi(q)&=&\int_{\mathcal{C}}dT\;\left\{\int_\text{b.c.}^{q(T)=q} \prod_{\tau=0}^T \left(dq(\tau)\;dp_q(\tau)\right)\;e^{+i\int_0^T d\tau\left(p_q\frac{dq}{d\tau}-H\right)}\right\},\\
&=&\int_{\mathcal{C}}dT\;\left\{\int_\text{b.c.}^{q(T)=q} \mathcal{D}q(\tau)\;e^{+iI[q(\tau),\Dot{q}(\tau);\;T]}\right\},\\
&=&\int_{\mathcal{C}}dT \mel{q}{e^{+i
T H}}{\text{b.c.}},\\
&=&\braket{q}{\text{b.c.}}_\text{dyn}.\label{overlapmini}
\eea
In other words, this is the point-particle propagator from $\ket{\text{b.c.}}$ to $\ket{q}$ in proper time $T$ via the Hamiltonian $H$, followed by integration over proper time. This last step is what gurantees that the state does not depend on any intrinsic property of the worldline, thus encoding the covariance of the theory. The boundary condition “$\text{b.c.}$" has the same meaning as in \eqref{overlap}.

What is left of what was originally the class of geometries $\mathcal{C}$ in the full theory, is now the proper-time contour in the minisuperspace ansatz. It tells us over what complex histories to sum over. Given such a contour in the complex-$T$ plane we can try to compute the integral by deforming the contour to pass through a saddle-point, provided we do not cross any poles or branch cuts. The position of the saddle-points in the complex-$T$ plane depends on the boundary condition $q(T)=q$ on $\Sigma$ (of course, after having fixed “b.c."). Let us analyze the general pattern of these saddle-points and let us take “b.c." to correspond to the no-boundary proposal for simplicity (although the same pattern follows if the boundary condition is of the form $\ket{\text{b.c.}}=\ket{g_0}$).

The full gravitational action takes the form:
\bea
I[\textbf{g}]=\frac{1}{16\pi G_N}\int_\mathcal{M}\varepsilon \;(\textbf{R}-2\Lambda)+ \frac{1}{8\pi G_N}\int_{\partial\mathcal{M}}\sigma \;K,
\eea
where $\varepsilon$ and $\sigma$ are the volume forms on $\mathcal{M}$ and $\partial\mathcal{M}$ respectively. If we flip the “time" orientation\footnote{Here “time" could be either Lorentzian or Euclidean time. In either case the orientation of the manifold would flip and so the conclusion follows. Also what matters is not that the boundary is spacelike but rather that it is orthogonal to the direction we are flipping.} of $\mathcal{M}$, then $\varepsilon\mapsto -\varepsilon$, while $\textbf{R}$ stays the same. Similarly, because any boundary of $\mathcal{M}$ is spacelike in the case at hand, we have that $K\mapsto -K$, while $\sigma$ remains unchanged. This means that under time-reversal the action flips sign:
\bea
\text{“Time" reversal}: I\mapsto -I.\label{timereversal}
\eea
Because a saddle-point of the functional $I[\textbf{g}]$ is also a saddle-point of $-I[\textbf{g}]$ and vice-versa, it follows that the saddle-points of the gravitational action come in “time"-reversed pairs. 

In the minisuperspace ansatz, the time-reversal transformation in encoded by the following relation:
\bea
I[q(\tau),\dot{q}(\tau);-T]=-I[q(\tau),\dot{q}(\tau);T], \;\;\;\forall T\in\mathbb{C}.\label{reversal}
\eea

Given our conventions, a value of $T\in \mathbb{R}$ corresponds to a Lorentzian geometry connecting the state $\ket{\text{b.c.}}$ to the state $\ket{q}$. $T>0$ means that $\ket{q}$ is to the future of $\ket{\text{b.c.}}$. Time-reversal would take $T$ to $-T$. Equation \eqref{reversal} tells us that a pair of such saddles would contribute with opposite phases to the path integral: $e^{+iI}$ and $e^{-iI}$. On the other hand, values of $T\in i\mathbb{R}$ correspond to geometries which are Euclidean (the action evaluated on them is purely imaginary). The sign of $T$ here means that either $\ket{q}$ or $\ket{\text{b.c.}}$ is to the “left" of the other. Equation \eqref{reversal} tells us that one of the saddles in the pair is exponentially enhanced/suppressed relative to the other. They would contribute to the path integral as: $e^{-|I|}$ and $e^{+|I|}$. 

We can also take the complex-conjugate of the action. Since $q(\tau)\in\mathbb{R}$ and in general $T\in\mathbb{C}$ we have that:
\bea
I[q(\tau),\dot{q}(\tau);T^*]=I^*[q(\tau),\dot{q}(\tau);T],\;\;\;\forall T\in\mathbb{C}.\label{cc}
\eea
But $I$ is a holomorphic function of $T$ (away from the essential singularity at $T=0$) and so if $T$ is a saddle-point of $I$, then $T$ is a saddle-point of $I^*$. Equation \eqref{cc} then implies that $T^*$ is also a saddle-point of $I$. Thus saddle-points come in complex-conjugate pairs.

\begin{figure}
    \centering

\tikzset{every picture/.style={line width=0.75pt}} %set default line width to 0.75pt        

\begin{tikzpicture}[x=0.75pt,y=0.75pt,yscale=-1,xscale=1]
%uncomment if require: \path (0,438); %set diagram left start at 0, and has height of 438

%Shape: Axis 2D [id:dp10757835097321311] 
\draw  (147,220.97) -- (488,220.97)(319.06,58) -- (319.06,379) (481,215.97) -- (488,220.97) -- (481,225.97) (314.06,65) -- (319.06,58) -- (324.06,65)  ;
%Straight Lines [id:da5087052011623879] 
\draw    (318.83,265.86) -- (319.27,310.78) ;
\draw [shift={(319.27,310.78)}, rotate = 89.44] [color={rgb, 255:red, 0; green, 0; blue, 0 }  ][fill={rgb, 255:red, 0; green, 0; blue, 0 }  ][line width=0.75]      (0, 0) circle [x radius= 3.35, y radius= 3.35]   ;
\draw [shift={(319.11,294.32)}, rotate = 269.44] [color={rgb, 255:red, 0; green, 0; blue, 0 }  ][line width=0.75]    (10.93,-3.29) .. controls (6.95,-1.4) and (3.31,-0.3) .. (0,0) .. controls (3.31,0.3) and (6.95,1.4) .. (10.93,3.29)   ;
%Straight Lines [id:da9264952446397221] 
\draw    (318.83,129.36) -- (318.91,174.18) ;
\draw [shift={(318.86,144.77)}, rotate = 89.89] [color={rgb, 255:red, 0; green, 0; blue, 0 }  ][line width=0.75]    (10.93,-3.29) .. controls (6.95,-1.4) and (3.31,-0.3) .. (0,0) .. controls (3.31,0.3) and (6.95,1.4) .. (10.93,3.29)   ;
\draw [shift={(318.83,129.36)}, rotate = 89.89] [color={rgb, 255:red, 0; green, 0; blue, 0 }  ][fill={rgb, 255:red, 0; green, 0; blue, 0 }  ][line width=0.75]      (0, 0) circle [x radius= 3.35, y radius= 3.35]   ;
%Straight Lines [id:da9021683118838106] 
\draw    (318.83,129.36) -- (409.87,129.66) ;
\draw [shift={(409.87,129.66)}, rotate = 0.19] [color={rgb, 255:red, 0; green, 0; blue, 0 }  ][fill={rgb, 255:red, 0; green, 0; blue, 0 }  ][line width=0.75]      (0, 0) circle [x radius= 3.35, y radius= 3.35]   ;
\draw [shift={(370.35,129.53)}, rotate = 180.19] [color={rgb, 255:red, 0; green, 0; blue, 0 }  ][line width=0.75]    (10.93,-3.29) .. controls (6.95,-1.4) and (3.31,-0.3) .. (0,0) .. controls (3.31,0.3) and (6.95,1.4) .. (10.93,3.29)   ;
%Straight Lines [id:da9732044403060574] 
\draw    (230.7,129.78) -- (318.83,129.36) ;
\draw [shift={(267.76,129.61)}, rotate = 359.73] [color={rgb, 255:red, 0; green, 0; blue, 0 }  ][line width=0.75]    (10.93,-3.29) .. controls (6.95,-1.4) and (3.31,-0.3) .. (0,0) .. controls (3.31,0.3) and (6.95,1.4) .. (10.93,3.29)   ;
\draw [shift={(230.7,129.78)}, rotate = 359.73] [color={rgb, 255:red, 0; green, 0; blue, 0 }  ][fill={rgb, 255:red, 0; green, 0; blue, 0 }  ][line width=0.75]      (0, 0) circle [x radius= 3.35, y radius= 3.35]   ;
%Straight Lines [id:da07731911445121586] 
\draw    (319.27,310.78) -- (410.91,310.4) ;
\draw [shift={(410.91,310.4)}, rotate = 359.76] [color={rgb, 255:red, 0; green, 0; blue, 0 }  ][fill={rgb, 255:red, 0; green, 0; blue, 0 }  ][line width=0.75]      (0, 0) circle [x radius= 3.35, y radius= 3.35]   ;
\draw [shift={(371.09,310.56)}, rotate = 179.76] [color={rgb, 255:red, 0; green, 0; blue, 0 }  ][line width=0.75]    (10.93,-3.29) .. controls (6.95,-1.4) and (3.31,-0.3) .. (0,0) .. controls (3.31,0.3) and (6.95,1.4) .. (10.93,3.29)   ;
%Straight Lines [id:da11572088803201208] 
\draw    (229.91,310.4) -- (319.27,310.78) ;
\draw [shift={(267.59,310.56)}, rotate = 0.24] [color={rgb, 255:red, 0; green, 0; blue, 0 }  ][line width=0.75]    (10.93,-3.29) .. controls (6.95,-1.4) and (3.31,-0.3) .. (0,0) .. controls (3.31,0.3) and (6.95,1.4) .. (10.93,3.29)   ;
\draw [shift={(229.91,310.4)}, rotate = 0.24] [color={rgb, 255:red, 0; green, 0; blue, 0 }  ][fill={rgb, 255:red, 0; green, 0; blue, 0 }  ][line width=0.75]      (0, 0) circle [x radius= 3.35, y radius= 3.35]   ;
%Straight Lines [id:da4960731746354681] 
\draw    (319.06,220.97) -- (318.91,174.18) ;
\draw [shift={(318.91,174.18)}, rotate = 269.82] [color={rgb, 255:red, 0; green, 0; blue, 0 }  ][fill={rgb, 255:red, 0; green, 0; blue, 0 }  ][line width=0.75]      (0, 0) circle [x radius= 3.35, y radius= 3.35]   ;
\draw [shift={(318.97,191.58)}, rotate = 89.82] [color={rgb, 255:red, 0; green, 0; blue, 0 }  ][line width=0.75]    (10.93,-3.29) .. controls (6.95,-1.4) and (3.31,-0.3) .. (0,0) .. controls (3.31,0.3) and (6.95,1.4) .. (10.93,3.29)   ;
\draw [shift={(319.06,220.97)}, rotate = 314.82] [color={rgb, 255:red, 0; green, 0; blue, 0 }  ][line width=0.75]    (-5.59,0) -- (5.59,0)(0,5.59) -- (0,-5.59)   ;
%Straight Lines [id:da4739700638529786] 
\draw    (319.06,220.97) -- (318.83,265.86) ;
\draw [shift={(318.83,265.86)}, rotate = 90.3] [color={rgb, 255:red, 0; green, 0; blue, 0 }  ][fill={rgb, 255:red, 0; green, 0; blue, 0 }  ][line width=0.75]      (0, 0) circle [x radius= 3.35, y radius= 3.35]   ;
\draw [shift={(318.92,249.41)}, rotate = 270.3] [color={rgb, 255:red, 0; green, 0; blue, 0 }  ][line width=0.75]    (10.93,-3.29) .. controls (6.95,-1.4) and (3.31,-0.3) .. (0,0) .. controls (3.31,0.3) and (6.95,1.4) .. (10.93,3.29)   ;
%Curve Lines [id:da8150311745805017] 
\draw  [dash pattern={on 0.84pt off 2.51pt}]  (381.05,257.59) .. controls (347.73,258.57) and (343.22,252.83) .. (327.05,230.95) ;
\draw [shift={(326.05,229.59)}, rotate = 53.53] [color={rgb, 255:red, 0; green, 0; blue, 0 }  ][line width=0.75]    (10.93,-3.29) .. controls (6.95,-1.4) and (3.31,-0.3) .. (0,0) .. controls (3.31,0.3) and (6.95,1.4) .. (10.93,3.29)   ;
%Curve Lines [id:da14872996817911366] 
\draw [color={rgb, 255:red, 208; green, 2; blue, 27 }  ,draw opacity=1 ]   (151.14,261.41) .. controls (180.2,231) and (299.95,230.41) .. (318.95,230.41) .. controls (337.95,230.41) and (459.2,230) .. (489.95,260.41) ;
\draw [shift={(237.74,234.63)}, rotate = 173.31] [fill={rgb, 255:red, 208; green, 2; blue, 27 }  ,fill opacity=1 ][line width=0.08]  [draw opacity=0] (8.93,-4.29) -- (0,0) -- (8.93,4.29) -- cycle    ;
\draw [shift={(412.12,235.37)}, rotate = 186.81] [fill={rgb, 255:red, 208; green, 2; blue, 27 }  ,fill opacity=1 ][line width=0.08]  [draw opacity=0] (8.93,-4.29) -- (0,0) -- (8.93,4.29) -- cycle    ;
%Curve Lines [id:da7552491150454903] 
\draw [color={rgb, 255:red, 208; green, 2; blue, 27 }  ,draw opacity=1 ]   (150,180) .. controls (180.2,210) and (300.2,210) .. (319.2,210) .. controls (338.2,210) and (460.2,210) .. (490.2,180) ;
\draw [shift={(237.71,206.14)}, rotate = 186.39] [fill={rgb, 255:red, 208; green, 2; blue, 27 }  ,fill opacity=1 ][line width=0.08]  [draw opacity=0] (8.93,-4.29) -- (0,0) -- (8.93,4.29) -- cycle    ;
\draw [shift={(412.67,204.87)}, rotate = 173.14] [fill={rgb, 255:red, 208; green, 2; blue, 27 }  ,fill opacity=1 ][line width=0.08]  [draw opacity=0] (8.93,-4.29) -- (0,0) -- (8.93,4.29) -- cycle    ;
%Shape: Arc [id:dp7380916533786287] 
\draw  [draw opacity=0] (289.06,220.97) .. controls (289.06,204.4) and (302.5,190.97) .. (319.06,190.97) .. controls (319.06,190.97) and (319.06,190.97) .. (319.06,190.97) -- (319.06,220.97) -- cycle ; \draw [color={rgb, 255:red, 208; green, 2; blue, 27 }  ,draw opacity=1 ]   (289.06,220.97) .. controls (289.06,204.4) and (302.5,190.97) .. (319.06,190.97) ;  
%Shape: Arc [id:dp9649444738295782] 
\draw  [draw opacity=0] (319.06,250.97) .. controls (319.06,250.97) and (319.06,250.97) .. (319.06,250.97) .. controls (302.5,250.97) and (289.06,237.54) .. (289.06,220.97) -- (319.06,220.97) -- cycle ; \draw [color={rgb, 255:red, 208; green, 2; blue, 27 }  ,draw opacity=1 ]   (319.06,250.97) .. controls (303.49,250.97) and (290.69,239.1) .. (289.21,223.92) ; \draw [shift={(289.06,220.97)}, rotate = 78.44] [fill={rgb, 255:red, 208; green, 2; blue, 27 }  ,fill opacity=1 ][line width=0.08]  [draw opacity=0] (8.93,-4.29) -- (0,0) -- (8.93,4.29) -- cycle    ; 
%Shape: Arc [id:dp7897074925285006] 
\draw  [draw opacity=0] (349.06,220.97) .. controls (349.06,220.97) and (349.06,220.97) .. (349.06,220.97) .. controls (349.06,237.54) and (335.63,250.97) .. (319.06,250.97) -- (319.06,220.97) -- cycle ; \draw [color={rgb, 255:red, 208; green, 2; blue, 27 }  ,draw opacity=1 ]   (349.06,220.97) .. controls (349.06,220.97) and (349.06,220.97) .. (349.06,220.97) .. controls (349.06,237.54) and (335.63,250.97) .. (319.06,250.97) ;  
%Shape: Arc [id:dp4276828026960776] 
\draw  [draw opacity=0] (319.06,190.97) .. controls (335.63,190.97) and (349.06,204.4) .. (349.06,220.97) -- (319.06,220.97) -- cycle ; \draw [color={rgb, 255:red, 208; green, 2; blue, 27 }  ,draw opacity=1 ]   (319.06,190.97) .. controls (334.64,190.97) and (347.44,202.84) .. (348.92,218.02) ; \draw [shift={(349.06,220.97)}, rotate = 258.44] [fill={rgb, 255:red, 208; green, 2; blue, 27 }  ,fill opacity=1 ][line width=0.08]  [draw opacity=0] (8.93,-4.29) -- (0,0) -- (8.93,4.29) -- cycle    ; 

% Text Node
\draw (301,33.4) node [anchor=north west][inner sep=0.75pt]    {$\text{Im}( T)$};
% Text Node
\draw (500,212.4) node [anchor=north west][inner sep=0.75pt]    {$\text{Re}( T)$};
% Text Node
\draw (296,99.4) node [anchor=north west][inner sep=0.75pt]    {$q=L_{\text{dS}}^{2}$};
% Text Node
\draw (259,161.4) node [anchor=north west][inner sep=0.75pt]    {$q< L_{\text{dS}}^{2}$};
% Text Node
\draw (423,115.4) node [anchor=north west][inner sep=0.75pt]    {$q >L_{\text{dS}}^{2}$};
% Text Node
\draw (169,115.4) node [anchor=north west][inner sep=0.75pt]    {$q >L_{\text{dS}}^{2}$};
% Text Node
\draw (384,248) node [anchor=north west][inner sep=0.75pt]   [align=left] { singularity};
% Text Node
\draw (129,162.4) node [anchor=north west][inner sep=0.75pt]    {$\textcolor[rgb]{0.82,0.01,0.11}{\mathcal{C}_{1}}$};
% Text Node
\draw (129,263.4) node [anchor=north west][inner sep=0.75pt]    {$\textcolor[rgb]{0.82,0.01,0.11}{\mathcal{C}_{2}}$};
% Text Node
\draw (283,245.4) node [anchor=north west][inner sep=0.75pt]    {$\textcolor[rgb]{0.82,0.01,0.11}{\mathcal{C}_{3}}$};

\end{tikzpicture}
    \caption{The dots represent the saddle-points of the action as the boundary condition $q$ increases across the cosmological scale, for the no-boundary proposal. For every saddle-point at $T$ there is another one at $-T$ and $T^*$. For $q<L_\text{dS}^2$ there are two saddle-points, both along the imaginary axis. These correspond to Euclidean geometries. For $q>L_\text{dS}^2$ there are four saddle-points, which correspond to complex geometries. The transition point happens at $q=L_\text{dS}^2$. In red are three contour examples. Each can be deformed to pass through a different set of saddle-points. There is an essential singularity at the origin.}
    \label{saddles}
    \end{figure}
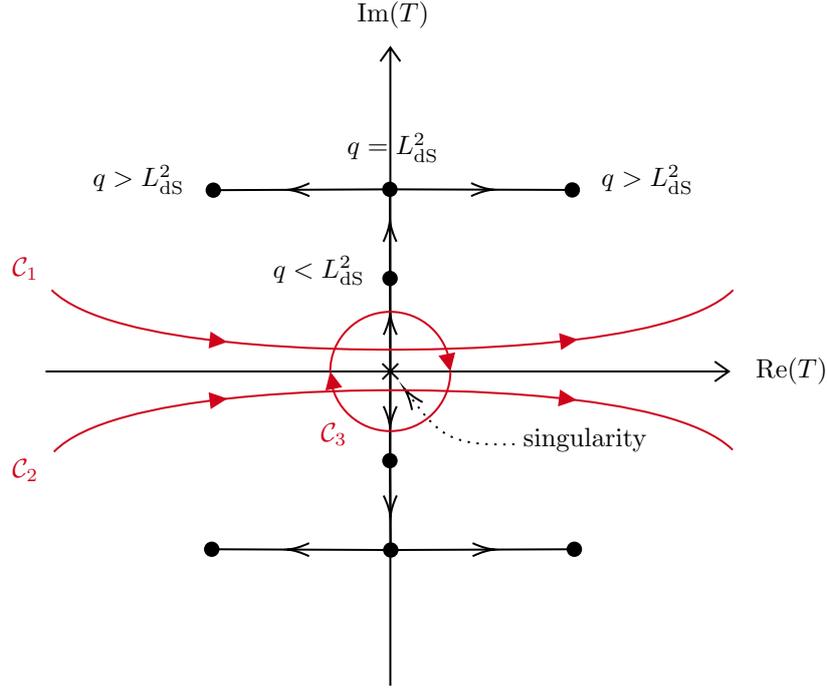

Figure \ref{saddles} highlights the general pattern of how, in minisuperspace, the saddle-points are distributed in the complex-$T$ plane for the case of the no-boundary proposal for $\ket{\text{b.c.}}$, in which $q(0)=0$. The story is analogous for any $\ket{\text{b.c.}}$ (unless the boundary condition explicitly breaks either time-reversal or complex-conjugate symmetry, as does for example $\ket{\text{b.c.}}=\ket{\Pi\neq0}$; in those cases, the analysis has to be performed again).

A general $T\in\mathbb{C}$ corresponds to a complex geometry interpolating between $\ket{\text{b.c.}}$ and $\ket{q}$. For any such geometry with $T=T_R +iT_I$ we can redefine the coordinate $t$ in the following way:
\bea
t \mapsto \Tilde{t}(t)=\begin{cases} 
      \frac{iT_I}{T}\;2t & 0\leq t\leq 1/2,\\
      \frac{T_R}{T}\;2\left(t-\frac{1}{2}\right) & 1/2\leq t\leq 1.
   \end{cases}
\eea
Then the evaluation of the action $I$ on such a complex geometry corresponds to splitting the action into two terms, one with purely Euclidean proper time and the other with purely Lorentzian proper time. For the no-boundary proposal, the complex saddle-points in Figure \ref{saddles} all have: $T=T_R\pm iL_\text{dS}$. So the action evaluated on such saddles splits as:
\bea
I[T]\vert_\text{N.B.}^q&=&I[\pm iL_\text{dS}]\vert_\text{N.B.}^{L_\text{dS}^2} + I[T_R]\vert_{L_\text{dS}^2}^q,\\
&=&\pm i \;I_E(L_\text{dS}^2)+I_L(q).
\eea
Here, $I_E(q)$ stands for the action evaluated on a purely Euclidean saddle geometry interpolating between the “no-boundary" and a slice with geometry given by $q$. Similarly, $I_L(q)$ stands for the action evaluated on a purely Lorentzian saddle geometry interpolating between a slice of area $L^2_\text{dS}$ and one of size $q$. This recovers the usual saddle-points considered by HH and Vilenkin in minisuperspace.

Now we are in a position to understand how different contours for $T$ can lead to different physical predictions. One crucial requirement for the contour is that it yields a quantum state satisfying the Hamiltonian constraint. In \cite{Teitelboim:1983fh} it was argued that a fundamental choice must be made in quantum gravity between preserving (what the author defined as) causality and gauge invariance. In particular, if we want to define a “causal" inner product $\braket{q_2}{q_1}_\text{dyn}$, such that we only consider histories in which the slice $\Sigma_2$ is to the future of $\Sigma_1$, we must choose the contour $\mathcal{C}$ to be on the positive real axis only, or in other words we only consider histories with positive lapse between the two slices. However, in such a theory, a quantum state of the form \eqref{overlapmini} with $\ket{\text{b.c.}}=\ket{q_0}$ only satisfies the Hamiltonian constraint up to a delta function:
\bea
H\Psi(q):=H\braket{q}{q_0}_\text{dyn}^\text{“causal"}=-i\delta(q-q_0). \label{causal}
\eea
The problematic point in configuration space is $q=q_0$ for which there are interpolating geometries with zero lapse. The Hamiltonian constraint ensures that the state is invariant under deforming the slice $\Sigma$ (on which the state lives) forwards or backwards in time. But for histories of zero lapse the choice of “causal" contour forbids infinitesimal deformations backwards in time. Thus, diffeomorphism invariance normal to $\Sigma$ is broken. For $q\neq q_0$ all interpolating geometries have non-zero lapse and so infinitesimal deformations are allowed in both directions. So the Hamiltonian constraint holds for those points in configuration space. 

Throughout this paper, we demand to preserve bulk gauge invariance and so we want the Hamiltonian constraint to be satisfied exactly. This means we are interested in quantum gravity theories defined by contours which allow deformations in both directions around any given lapse. Therefore, in this paper, we only consider contours with \emph{no endpoints}.\footnote{Actually, at the quantum level, what we really need is that the lapse integral in the gravitational path integral \eqref{state} provides a representation of $\delta(H)$ so that the Hamiltonian constraint is imposed at the final slice. In other words, after gauge-fixing, we need the lapse integral to be in some sense isomorphic to:
\bea
\int_{-\infty}^{+\infty}dN\;e^{-iNH}\sim \delta(H).
\eea
The contours $\mathcal{C}_{1/2}$ can be deformed to contours parallel to the real axis and so indeed impose the Hamiltonian constraint. The contour $\mathcal{C}_3$ is a superposition of the previous two and so also provides a good representation. It would be interesting to understand which contours, beyond these three examples, impose bulk diffeomorphism invariance in this path integral sense.}
Three such examples are provided in Figure \ref{saddles}. These are guaranteed to give quantum states satisfying the Hamiltonian constraint.

The contours shown can be deformed to pass through the appropriate saddle-points along steepest descent directions. We can then evaluate the path integral to give, schematically:
\bea
\mathcal{C}_1 \rightarrow \Psi_1(q)\sim\begin{cases} 
      e^{-|I_E(q)|} & 0< q< L^2_\text{dS},\\
      e^{-|I_E(L^2_\text{dS})|}\left(e^{+iI_L(q)}+e^{i\varphi_1}e^{-iI_L(q)}\right) & L^2_\text{dS}<q.
   \end{cases}\label{contour1}
\eea
\bea
\mathcal{C}_2 \rightarrow \Psi_2(q)\sim \begin{cases} 
      e^{+|I_E(q)|} & 0< q< L^2_\text{dS},\\
      e^{+|I_E(L^2_\text{dS})|}\left(e^{+iI_L(q)}+e^{i\varphi_2}e^{-iI_L(q)}\right) & L^2_\text{dS}<q.
   \end{cases}\label{contour2}
\eea
The time-reversal and complex-conjugate transformation properties of the action discussed earlier lead to the contributions above. In the classically allowed region of $q>L^2_\text{dS}$ we get oscillatory behaviour as we vary $q$. We have left the relative phases $(\varphi_1,\varphi_2)$ unspecified. Determining them requires a full Piccard-Lefschetz analysis of the path integral. For this we refer the reader to \cite{Feldbrugge:2017kzv}. The result (explained in that paper) is that the oscillatory behaviour of the quantum state for $q>L^2_\text{dS}$ is $\pi/2$ out of phase for the two cases of contour. We further note that in the classically forbidden region of $0<q<L^2_\text{dS}$, the contours $\mathcal{C}_{1/2}$ give an exponential suppression/enhancement, respectively, as can be seen in \eqref{contour1} and \eqref{contour2}.\footnote{We need to normalize the results by factors of $e^{-I_\text{sphere}}:=e^{-2|I_E(L^2_\text{dS})|}$ in order to get actually what the superposition of field theory branches gives in the Euclidean region. This factor is independent of $q$ and so is just an overall normalization.}

By comparing with the behaviour of the solutions obtained in Section \ref{mini}, we can hypothesize the following identification to be valid even beyond the saddle-point approximation of the path integral:
\bea
\mathcal{C}_1 &\leftrightarrow& \Psi_\text{HH}=\frac{Z_++Z_-}{2},\\
\mathcal{C}_2 &\leftrightarrow& \Psi_{\overline{\text{HH}}}=\frac{Z_+-Z_-}{2i},\\
\mathcal{C}_3 &\leftrightarrow& \Psi_\text{HH}-\Psi_{\overline{\text{HH}}}=\frac{1+i}{2}Z_++\frac{1-i}{2}Z_-.
\eea
From the perspective of Cauchy Slice Holography, the only difference between all these sates is the linear combination of the branches $(Z_\pm)$ that we chose. This minisuperspace analysis thus suggests that the choice of branch superposition is somehow in close correspondence with the choice of lapse contour in the gravitational path integral. 

To make the connection sharp, we would need to properly study the space of all allowed contours and how the saddle-points contribute via Piccard-Lefschetz theory. For example, it would be interesting to see what contours would lead to differently weighted superpositions of $\Psi_\text{HH}$ and $\Psi_{\overline{\text{HH}}}$. We leave this for future work. 

There has been recent progress in the literature regarding what the physically sensible contours to use in the gravitational path integral are. This is based on the idea that if we were to couple QFT to a given geometry that we happened to decide to sum over, then the QFT path integral on this background should be convergent \cite{Witten:2021nzp}. Based on this criterion several regions of the complex-$T$ plane turn out to be forbidden \cite{Lehners:2021mah, Jonas:2022uqb}. It would be interesting to understand whether this story is consistent with the quantum states being defined by superpositions of $(Z_\pm)$.

\section{Comments on the Structure of Quantum Gravity}\label{CPT}

We now turn attention to the relationship between various properties of the bulk and dual theories. We should stress that all that follows will be done for pure gravity. We leave for future work how the inclusion of matter might modify the discussion.

We discuss how the two field theory branches are related by a CPT transformation and how imposing bulk CPT symmetry restricts the choice of class of geometries $\mathcal{C}$ in \eqref{dynamical}. Then, we present an argument for bulk unitarity in full quantum gravity. In the context of AdS/CFT, bulk unitarity can be understood to follow from boundary unitarity. However, in the case of closed cosmology, the dual Euclidean field theory turns out to \emph{not} be reflection-positive, as we explicitly show in the case corresponding to a CPT-invariant bulk. We end by discussing, following our analysis, the possibility that the Hilbert space of quantum gravity is one-dimensional.

\subsection{CPT and Quantum Gravity}

In QFT, the CPT-theorem states that a unitary, locally Lorentz-invariant field theory is invariant under a CPT transformation. The converse is not generically true. It is not clear if such a theorem exists for quantum gravity. We definitely want our quantum gravity theory to be unitary (a fact we will show shortly) and the simplest way to make CPT a symmetry of matter in Nature would seem to be to impose it at the quantum gravitational level.\footnote{It could be the case that the CPT-theorem only holds in an approximate sense in the limit of QFT in curved spacetime and that it is violated if we include quantum gravity effects. At the moment we do not see any evidence in favour of this possibility.} But we will see, when we discuss bulk unitarity, that imposing CPT-symmetry seems to be an independent statement, thus suggesting that an analogue of the CPT-theorem does \emph{not} hold in quantum gravity. Nevertheless, it is worthwhile understanding what it would mean to have a CPT-invariant theory of quantum gravity. 

Imposing CPT invariance of the theory means taking a class of histories $\mathcal{C}$ in \eqref{dynamical} such that any result of the path integral computation is CPT-invariant, unless this is explicitly broken by the boundary conditions. In particular, this means that a quantum gravity state $\Psi[g]=\braket{g}{\text{b.c.}}_\text{dyn}$ with $\ket{\text{b.c.}}=\ket{g_0}$ must be CPT-invariant. (A boundary condition like $\ket{\text{b.c.}}=\ket{\Pi\neq 0}$ would break CPT explicitly.) In the pure gravity minisuperspace toy model, where the class $\mathcal{C}$ corresponds to the set of lapse contours, imposing CPT-invariance corresponds to choosing contours that are invariant under reversal of Lorentzian time, i.e. invariant under reflection about the imaginary lapse axis. All three examples shown in Figure \ref{saddles} satisfy that property.

But what does a CPT transformation look like from the perspective of a quantum state living on $\Sigma$? C and P act as expected. Focusing on $d=2$ and trivial topology for concreteness, P takes $S^2$ to itself and C acts trivially when there is no matter.\footnote{Actually, at the level of analysis in this paper all we can really talk about is the relationship between PT and unitarity, since we are not including any matter, thus making the C transformation meaningless. However, there are all the reasons to believe that CPT is the correct transformation to consider when matter is included. But we leave all that matter for future work.} This has no effect on covariant integrals on $S^2$ and so leaves the quantum state invariant. But the state $\Psi$ encodes information both about “position" and “velocity". In the “position" representation, the “velocity" is given by the operator $\sim i \partial/\partial g_{ab}$. Thus, to reverse the “velocity", as the T operation must do, we must complex conjugate. Hence we have:
\bea
\text{T}:i\mapsto-i.
\eea
This is consistent with T being an antilinear map. It follows that the combined action on states in the Hilbert space corresponding to pure gravity is:
\bea
\text{CPT}:\Psi\mapsto\Psi^*.
\eea
At the level of operators the action is:
\bea
\text{CPT}: \;\;g_{ab}(x)&\mapsto& g_{ab}(\hat{\mathcal{P}}x),\\
\Pi^{ab}(x)&\mapsto& - \Pi^{ab}(\hat{\mathcal{P}}x),
\eea
where $\hat{\mathcal{P}}x$ is the reflected point corresponding to $x$. In $d=2$ spatial dimensions a parity transformation is given by a reflection about a one-dimensional cut. 
The Hamiltonian constraint operator is CPT-invariant in the sense that:
\bea
\text{CPT}:\;\;\mathcal{H}(x)\mapsto\mathcal{H}(\hat{\mathcal{P}}x),
\eea
since all the coefficients are real and it is quadratic in $\Pi^{ab}$. The momentum constraint operator is invariant in the same sense. The fact that the Hamiltonian constraint is CPT-invariant also means that if $\Psi$ is a solution then so is $\text{CPT}(\Psi)=\Psi^*$. This is reflected in the fact that the two branches $Z_\pm$ are complex conjugates of each other, as was explicitly shown in the minisuperspace model. This property must hold even when all the gravitational  degrees of freedom are turned on. When matter is included the two branches $Z_\pm$ will still be CPT duals of each other, however in that case a CPT transformation would be more than just complex conjugation.

Demanding CPT invariance of the quantum state \eqref{superposition} reduces the space of possible solutions to the Hamiltonian constraint to the following subset:
\bea
\Psi[g]=A\left(e^{+i\varphi}Z_+[g]+e^{-i\varphi}Z_-[g]\right)=\Psi^*[g],\label{CPTsubspace}
\eea
where $A,\varphi\in\mathbb{R}$. In the language of Section \ref{contour}, we can equivalently say that choosing a class of histories $\mathcal{C}$ that gives a CPT-invariant theory corresponds to considering only a subset of possible branch superpositions. The two minisuperspace examples plotted in Section \ref{mini} correspond to $\varphi=0$ and $\varphi=\pi/2$ respectively. The significance of the normalization factor $A$ will be discussed later. The Vilenkin state \eqref{Vilenkin} is not real and so not CPT-invariant. It describes a tunnelling process from nothing into a future-directed expanding Universe.\footnote{It is defined via a path intgral over positive real lapse only. This means it satisfies the Hamiltonian constraint everywhere in the range $0<q<\infty$, but not at $q=0$, as discussed in the previous section.}

What follows does not rely on imposing bulk CPT symmetry. However, for purposes of illustration, we will sometimes restrict to that case where things are clearer.

%In the spirit of the previous section, a CPT-invariant state would be associated to a CPT-invariant class of histories $\mathcal{C}$. In the minisuperspace model, we made an identification between certain lapse contours and certain quantum states, which all were of the form \eqref{CPTsubspace}. Indeed, by looking at Figure \ref{saddles} we can see that all three contours map to themselves under time-reversal so the result of the path integral remains unchanged under a CPT transformation.

%In order to be able to construct bulk CPT invariant states it was crucial that the two deformed field theories were CPT duals of each other, otherwise whatever phase relation we imposed for a given $|u|$ would be spoiled when $|u|$ is varied. 

\subsection{Breakdown of Reflection Positivity}

In minisuperspace, we saw that states of the form \eqref{CPTsubspace} oscillate in the classically allowed region. This oscillatory nature has a striking consequence from the perspective of the field theory, namely the fact that it is a Euclidean field theory which does not satisfy reflection-positivity. We will argue for this in minisuperspace, but we expect it to be true in the full theory. Also we will only explicitly show it in the case of a CPT-invariant bulk, but things are even worse if we relax that.

Whatever linear combination of $Z_\pm$ we consider, we can interpret it as itself a partition function of some field theory living on $S^2$:
\bea
Z[g]:= A_+ Z_+[g]+A_-Z_-[g]
\eea
In the CPT-invariant subspace of quantum gravity theories we have that $Z[g]=\left(Z[g]\right)^*$.

\begin{figure}[H]
    \centering
\tikzset{every picture/.style={line width=0.75pt}} %set default line width to 0.75pt        

\begin{tikzpicture}[x=0.75pt,y=0.75pt,yscale=-0.75,xscale=0.75]

%Shape: Circle [id:dp8138756587243764] 
\draw   (219,151.5) .. controls (219,96) and (264,51) .. (319.5,51) .. controls (375,51) and (420,96) .. (420,151.5) .. controls (420,207) and (375,252) .. (319.5,252) .. controls (264,252) and (219,207) .. (219,151.5) -- cycle ;
%Shape: Arc [id:dp36331541516176236] 
\draw  [draw opacity=0] (313.97,252) .. controls (313.96,252) and (313.95,252) .. (313.94,252) .. controls (297.37,252) and (283.94,207.23) .. (283.94,152) .. controls (283.94,96.77) and (297.37,52) .. (313.94,52) .. controls (313.96,52) and (313.97,52) .. (313.99,52) -- (313.94,152) -- cycle ; \draw   (313.97,252) .. controls (313.96,252) and (313.95,252) .. (313.94,252) .. controls (297.37,252) and (283.94,207.23) .. (283.94,152) .. controls (283.94,96.77) and (297.37,52) .. (313.94,52) .. controls (313.96,52) and (313.97,52) .. (313.99,52) ;  
%Shape: Arc [id:dp7245894885059917] 
\draw  [draw opacity=0][dash pattern={on 4.5pt off 4.5pt}] (313.01,51.08) .. controls (313.01,51.08) and (313.02,51.08) .. (313.02,51.08) .. controls (329.59,50.93) and (343.43,95.79) .. (343.93,151.27) .. controls (344.44,206.55) and (331.51,251.52) .. (315.02,251.99) -- (313.94,151.54) -- cycle ; \draw  [dash pattern={on 4.5pt off 4.5pt}] (313.01,51.08) .. controls (313.01,51.08) and (313.02,51.08) .. (313.02,51.08) .. controls (329.59,50.93) and (343.43,95.79) .. (343.93,151.27) .. controls (344.44,206.55) and (331.51,251.52) .. (315.02,251.99) ;  

% Text Node
\draw (246,142.4) node [anchor=north west][inner sep=0.75pt]    {$\bra{\psi }$};
% Text Node
\draw (293,142.4) node [anchor=north west][inner sep=0.75pt]    {$\ket{\psi }$};

\end{tikzpicture}
    \caption{Symmetric cut of $S^2$ allows for reinterpretation of the partition function as a norm $\braket{\psi}$ for some state $\ket{\psi}$ living on the cut.}
    \label{RP}
\end{figure}
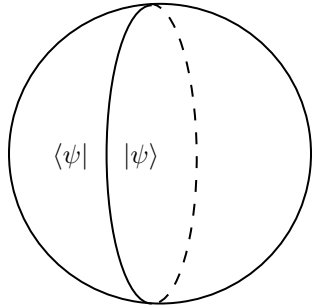

Restricting now to minisuperspace (so we have $Z[g]=Z(u)$), if we consider some foliation of this $S^2$ spatial slice, and assuming this theory satisfies the usual properties of Euclidean field theory, then we will have Hilbert spaces living on those foliations. Consider one such symmetric cut, as shown in Figure \ref{RP}. We can think of this as gluing two path integrals on half spheres:
\bea
Z(u)&=&\int \mathcal{D}\chi \;Z^\text{left}(u)[\chi]\times Z^\text{right}(u)[\chi],\\
&=& \int \mathcal{D}\chi \;\left(Z^\text{right}(u)[\chi]\right)^*\times Z^\text{right}(u)[\chi],\\
&=&\braket{\psi}\in \mathbb{R},
\eea
for some state $\braket{\chi}{\psi}=Z^\text{right}(u)[\chi]$, where $\{\chi\}$ represents some boundary condition on the cut for the path integral. In the second line we used CPT invariance of the field theory partition function (CPT\footnote{Note that here we are talking about a CPT transformation of the dual field theory, and not of the bulk. Because the field theory is Euclidean, we need both P and T combined to reflect the slice $\Sigma$, since in this case T reflects Euclidean time. Still, it is an antilinear operator and so it takes $i\mapsto -i$. So, in this subtle manner, both from the perspectove of the dual field theory and from the perspective of canonical quantization of gravity, CPT means the same thing operationally.} both complex conjugates and flips the left half sphere onto the right half sphere; in minisuperspace the background geometry is also the same on either side of the cut). The CPT-conjugate of the partition function creates the bra state.

Thus we can interpret the partition function on $S^2$ as the norm of a state in this field theory. Because $Z(u)$ oscillates, as shown in the explicit examples in Section \ref{mini}, this directly proves that there are many states $\ket{\psi}$, corresponding to different background values $u$, which have \emph{negative norm}. In other words, reflection-positivity does not hold for such field theories.

We can understand this also in the following way. In usual Euclidean QFT, reflection-positivity implies CPT symmetry. However, even though $Z$ is CPT-invariant, it \emph{spontaneously} breaks it. This is because it is a sum of two terms ($Z_\pm$) which individually are not CPT-invariant, as can be checked explicitly from \eqref{solution}. So we expect a breakdown of reflection-positivity for $Z$. Obviously, this also means that none of the branches is independently reflection-positive. It can also be checked that in the limit $u\to 0$ each branch recovers individual CPT-invariance. So we might wonder whether reflection-positivity could in principle be recovered in the deep UV. This minisuperspace analysis leads us to conjecture that, in full quantum gravity, whatever the UV completion of $Z[g]$ might be, it should have the property that CPT-symmetry is spontaneously broken at some point along its RG flow.

\subsection{Bulk Dynamical Positivity} \label{unitarity}

Nevertheless, this is still consistent with a positive-definite bulk \emph{dynamical} inner product for \emph{closed} quantum cosmology, as we will argue now.

It is worth noting that, until we specify the class of geometries $\mathcal{C}$ in \eqref{dynamical}, there is no universal notion of time-ordering between the two slices $\Sigma_{1/2}$. For example, if we take a $\mathcal{C}$ that only includes histories with positive lapse, then we can adopt a consistent convention such that $\Sigma_1$ is to be thought of as to the future of $\Sigma_2$.\footnote{This is what is done in \cite{Witten:2001kn} where an inner product is defined with reference to some hypothetical asymptotic future and past regions $\mathcal{I}_\pm$.} However, if we allow histories with both positive and negative lapse, then any notion of time-ordering cannot be made consistent across all histories (and even within a single history, since the sign of the lapse can vary spatially). Thus, any potential notion of time-ordering is to be thought of as emergent from $\mathcal{C}$ and in many cases it does not exist.\footnote{Given our discussion around \eqref{causal}, we should only consider contours which do not have endpoints and thus always include both signs of the lapse. So it seems that imposing a universal time-ordering is even inconsistent with diffeomorphism invariance.} In conclusion, we should remain agnostic about this point. 

Let us consider a hypothetical complex geometry, \textbf{g}, interpolating between two slices and thus contributing  to $\braket{g_1}{g_2}_\text{dyn}$. The action evaluated on it is complex. But as we discussed in Section \ref{contour}, it is always the case that a change of coordinates can be performed such that this geometry can be turned into a purely Euclidean piece glued to a purely Lorentzian piece. The action then can be written as:
\bea
I[\textbf{g}]=I_E+I_L,\;\;\; I_E\in i\mathbb{R},\;I_L\in\mathbb{R}.
\eea
Its contribution to the amplitude is:
\bea
e^{+iI[\textbf{g}]}=e^{+iI_L}e^{\mp|I_E|}.
\eea
If we complex conjugate this we get:
\bea
\left(e^{+iI[\textbf{g}]}\right)^*=e^{-iI_L}e^{\mp|I_E|},\label{conjugate}
\eea
showing that the Euclidean piece stays the same, while the Lorentzian one flips sign. But this is exactly what would happen if we decided to reverse the direction of Lorentzian time. This flips the time-ordering of the two slices, which means that \eqref{conjugate} is a contribution to the amplitude $\braket{g_2}{g_1}_\text{dyn}$, with bra and ket swapped. Because this property holds for any interpolating geometry we conclude that:
\bea
\braket{g_1}{g_2}^*_\text{dyn}=\braket{g_2}{g_1}_\text{dyn}.\label{hermitian}
\eea
The transition amplitude is thus Hermitian. Notice that this holds for any choice of $\mathcal{C}$, since it is true history by history.

But we argued in Section \ref{contour} that the overlap $\braket{g}{\text{b.c.}}_\text{dyn}$ defines a quantum state, for some boundary condition “b.c.". Looking at \eqref{hermitian} we get that:
\bea
\Psi_{g_2}^*[g_1]=\Psi_{g_1}[g_2],\label{equality}
\eea
where $\Psi_{g_2}$ is a quantum state defined by $\ket{\text{b.c.}}=\ket{g_2}$, while $\Psi_{g_1}$ is a quantum state defined by $\ket{\text{b.c.}}=\ket{g_1}$. The former satisfies the Hamiltonian constraint with respect to the variable $g_1$, while the latter does so with respect to $g_2$. This implies that each of them can be written in terms of a field theory partition function (defined via whichever branch superposition corresponds to the class $\mathcal{C}$ we happen to be considering). Thus, up to an overall normalization, we can say:
\bea
\Psi_{g_2}[g_1]&\sim& Z[g_1],\\
\Psi_{g_1}[g_2]&\sim& Z[g_2].
\eea
Comparing with \eqref{equality} we reach the conclusion that a quantum state $\Psi_{g_0}[g]=\braket{g}{g_0}_\text{dyn}$ has a holographic description given by:
\bea
\Psi_{g_0}[g]=Z^*[g_0]\times Z[g].\label{normalization}
\eea
We are led to the hypothesis that for quantum gravity on closed spatial slices $\Sigma$ the information about the boundary condition is provided by the relative normalization factors between states.\footnote{We leave for future work the interpretation of this in light of the problem of quantum mechanics applied to closed systems \cite{Hartle:1992as}. Should they really be interpreted as different states?}

For the transition amplitude \eqref{dynamical} this means the following:
\bea
\braket{g_1}{g_2}_\text{dyn}=Z[g_1]\times Z^*[g_2],\label{holography}
\eea
which is consistent with \eqref{hermitian}.
This equality turns out to be a special case of the Generalized Holographic Principle motivated in \cite{Araujo-Regado:2022gvw}. However, for the argument we want to make here we do not need to invoke it.

Dynamical positivity now follows trivially since:
\bea
||g||_\text{dyn}^2:=\braket{g}{g}_\text{dyn}=|Z[g]|^2\geq 0,\;\;\forall g.
\eea
Hence, the transition amplitude defined in \eqref{dynamical} is Hermitian and positive-semidefinite and so \emph{is} an inner product (after quotienting by null states), regardless of the choice of $\mathcal{C}$.\footnote{Had we been working with a CPT-invariant class $\mathcal{C}$, we would have had that $Z[g]\in\mathbb{R}$. In that case, the inner product \eqref{holography} has the further property of being real.} We call it the \emph{dynamical} inner product. We have just shown that we have \emph{bulk unitarity}.

Of course, we can make the inner product positive-definite by removing the null states, which exist as a consequence of the diffeomorphism invariance of the theory. In minisuperspace it is clear what to do, so we use it as an example of the quotient procedure, although the general message holds in full quantum gravity. We take a CPT-invariant theory as an example, since we have plotted solutions for such cases in Section \ref{mini}. The explicit solutions (and indeed any CPT-invariant solution) constructed in there oscillate around zero. Different branch superpositions will have zeroes at different points in configuration space. This is in line with the different superpositions corresponding to different theories, with different dynamical inner products.

We define the space of distributions on minisuperspace as (a much more detailed and less naive discussion about these distribution spaces is provided in \cite{Araujo-Regado:2022gvw}):
\bea
K^*:=\text{span}_\mathbb{C}\;\{\delta(q-\Tilde{q}): \Tilde{q}\in\mathbb{R}_+\}.
\eea
This is the vector space dual to the space of functions on minisuperspace. Equation \eqref{holography} defines an inner product on $K^*$.
Thus, we see that there is a non-trivial \emph{null} subspace of $K^*$, call it $K^*_0$, whose elements have zero dynamical inner product with everything in $K^*$:
\bea
K^*_0\supseteq\text{span}_\mathbb{C}\;\left\{\delta(q-\Tilde{q})\in K^*:Z(\Tilde{u})=0,\text{ for } \Tilde{u}=-\frac{\Tilde{q}}{\mu}\right\}\neq\emptyset.
\eea
Exactly which elements belong to $K^*_0$ depends on what quantum gravity theory we have, or equivalently, on the given linear combination defining $Z$.
The quotient space
\bea
Q:=K^*/K^*_0
\eea
therefore has a positive-definite inner product given by \eqref{dynamical} (or \eqref{holography}), by construction. 

A further interesting observation is that, since the argument above is independent of $\mathcal{C}$, we could have a quantum gravity theory that is unitary but not CPT-symmetric (by picking a $\mathcal{C}$ that explicitly breaks it). This seems to go against the intuition provided by the CPT-theorem in regular QFT. Does the QFT in curved spacetime statement emerge in the semiclassical limit of quantum gravity for any class $\mathcal{C}$? Or is the fact that we observe CPT symmetry for matter in Nature a hint that we should only consider CPT-invariant quantum gravity theories, thus restricting our options for $\mathcal{C}$? In order to tackle these important questions, we would need to first include matter and see what the necessary conditions are to have bulk unitarity in such a quantum gravity $+$ matter system. We leave this for future work.

The moral of the story is that even though the boundary field theory breaks reflection-positivity, the bulk quantum gravity theory is still unitary. 

\subsection{Holographic Sectors or Uniqueness?}\label{sectors}

However, perhaps the most crucial observation to make of the inner product \eqref{holography} is not that it is positive-definite, but rather that it is the natural inner product on the vector space $\mathbb{C}$. Thus, it is natural to identify:
\bea
Q\cong\mathbb{C},
\eea
as Hilbert spaces, where $Q$ is defined to be the space of distributions on configuration space (after quotieting), even beyond minisuperspace. It was argued in \cite{Araujo-Regado:2022gvw} that this space is isomorphic as a Hilbert space to the space of physical states, i.e. the space of solutions to the Hamiltonian constraint. Thus,
\bea
H_\text{phys}\cong \mathbb{C},\label{unique}
\eea 
which is a one-dimensional Hilbert space. 

Let us summarize the steps taken in reaching this seemingly drastic conclusion:
\begin{enumerate}
    \item Let us assume that we find a holographic field theory partition function (i.e. one that solves the Hamiltonian constraint) and we choose a particular branch superposition, call it $Z[g]$;
    \item This defines an inner product on the space of ``metric eigenstates", given by \eqref{holography}. These are the transition amplitudes of the quantum gravity theory which is holographically dual to our choice of $Z[g]$;
    \item In that specific quantum gravity theory, the physical Hilbert space of the constrained system on closed spatial slices is isomorphic to $\mathbb{C}$.
\end{enumerate}

Of course, this should not come entirely as a surprise. Another way to interpret the analysis is that by picking a particular $Z_\text{CFT}$ (and a particular linear superposition of branches) we are providing initial data for the Wheeler-DeWitt equation, which is a second-order hyperbolic equation on configuration space. The fact that this allows us to reconstruct a unique solution can be hardly unexpected.\footnote{One needs to be careful about the Cauchy problem in configuration space. Does the large volume limit correspond to a ``Cauchy" slice in configuration space? Relatedly, is the configuration space globally hyperbolic? See \cite{DeWitt:1967yk} for such discussions.} So one could push back by pointing out that all we have done is look at a one-dimensional sub-Hilbert space of the full space of solutions to the Hamiltonian constraint.

However, there are a few problems with such a stance: (We ignore branch superposition in the following arguments, for simplicity of the discussion)

\paragraph{Lack of Vector Space Structure}
Firstly, suppose we are provided with two solutions $\Psi_1$ and $\Psi_2$, corresponding to two different CFT partition functions in the large volume limit, $Z_{\text{CFT}1}$ and $Z_{\text{CFT}2}$ respectively. If both $\Psi_1, \Psi_2\in H_\text{phys}$ (where $H_\text{phys}$ is a hypothetical ``big" Hilbert space of quantum gravity), then we can superpose these solutions: $\Psi_1+\Psi_2$ to get a new solution. But what does it mean to have a vector space structure on the space of CFT partition functions? According to Freidel's analyisis, $\Psi_1+\Psi_2$ should also correspond to a CFT partition function in the large volume limit. Because generically two different CFTs will have wildly different spectra of operators, we do not believe such a summation makes sense. Thus, this would be an argument against the existence of such a ``big" vector space $H_\text{phys}$. 

More consistently, we could say that each Wheeler-DeWitt state corresponds to a quantum gravity sector, in other words to one (of the potentially many) allowed quantum gravity theories in the bulk. Each particular sector would correspond to a given holographic dual field theory, labelled by a given CFT in theory space. We thus reach a picture of ``holographic sectors" in quantum gravity. The sectors are to be thought of as having different kinds of objects, rendering meaningless a sum across sectors. 

\paragraph{Constraints on CFT Spectrum}
Secondly, the CFT behaviour only fixes the dependence on the Weyl factor of the spatial metric, but to fix the dependence on the other degrees of freedom of the metric we need to go to the next order in the equations. We have done this by performing the $T^2$ deformation of the CFT. But in order for this to make sense from the field theory side, the following operator must exist in the spectrum of the CFT:
\bea
:\left(\Pi_{ab}\Pi^{ab}-\frac{1}{d-1}\Pi^2\right):\label{protected}
\eea
and, crucially, it must have scaling dimension \emph{exactly} $2d$. However, such a dimension is not protected in a general interacting CFT. This is another way of seeing that it is difficult to find holographic duals to quantum gravity. 

Thus, it could be the case that there are not many partition functions $Z_\text{CFT}$ to choose from in the first place. Maybe there is only one, as the last point will try to illustrate. In the language of the previous paragraph, it could be the case that there is a \emph{unique} holographic sector. In other words, it could be the case that there is only one consistent theory of quantum gravity that can be written down. The holy-grail question still remains of what the ``correct" $Z_\text{CFT}$ is, however.

\paragraph{Analogy with AdS/CFT}
Lastly, the analogue scenario in AdS/CFT is the Wick-rotated picture of Wheeler-DeWitt states on radial slices of the bulk. In the large volume limit, all such states tend to CFT partition functions, consistent with the original duality. Also in that case, we could choose different starting CFTs on this radial boundary and obtain, correspondingly, different bulk radial states. However, we do not do that in AdS/CFT. We say that the full quantum gravity theory in the bulk is entirely described by \emph{a single} CFT partition function (including all its sources we could turn on). And had we found another candidate holographic CFT, we would say that it would be dual to a \emph{different} quantum gravity theory in the bulk. We certainly would not think of summing two such partition functions. In addition, in AdS/CFT we also believe holographic CFTs to be very special kinds of theories and it is not clear if, given a dimension and bulk matter content, more than one dual field theory could even be constructed. In our language, this is akin to the comment made above about the requirement to have a ``protected" $T^2$ operator in the spectrum. 

If we take all of this AdS/CFT intuition seriously and we translate the consequences to the analogous problem of $\Lambda>0$ on closed slices, we reach the picture advocated for above, which is neatly summarized by \eqref{unique}, namely that of a potentially unique holographic sector, up to the choice of branch superposition between the CPT-dual CFT partition functions. As explained earlier, this goes into the specification of the path integral contour in the bulk.

\section{Discussion}\label{discussion}

In this paper we explored how the framework of Cauchy Slice Holography applies to cosmology on closed spatial slices $\Sigma$. After explaining the prescription for the case at hand of $2+1$-dimensions, we calculated the quantum gravity state explicitly in minisuperspace. We obtained the space of solutions to the Hamiltonian constraint, which includes the familiar solutions of Hartle and Hawking and Vilenkin, by deforming two CFT branches. The deformation required is uniquely determined by the Hamiltonian operator. The correct CFT to start with could in principle be determined from IR properties of gravity. However, in the minisuperspace ansatz, the CFT functional dependence on the scale factor is fixed by the anomaly constraint and considerable progress can be made. Here we restricted attention to pure gravity, although the prescription could be easily extended to include matter. In particular, it would be interesting to follow the same steps for a minimally coupled scalar field in minisuperspace, as this might give insight into inflationary models.

\paragraph{Branch Superposition}

The most important equation in the paper is \eqref{superposition}:
\bea
\Psi[g]=A_+Z_+[g]+A_-Z_-[g]=:Z[g],\label{branches}
\eea
which encodes the fact that a quantum gravity state is a superposition of two branches. The partition functions on each branch are complex conjugates of each other, as we showed explicitly in minisuperspace. In fact, they are CPT-duals of each other, which is equivalent to complex conjugation for pure gravity. A quantum gravity state $\Psi[g]$ lives on an abstract slice $\Sigma$. But through each branch it has information about both future and past directions relative to $\Sigma$. If we were to define the state via a gravitational path integral, then in the saddle-point approximation, each branch would correspond to saddle histories of opposite time orientations. By picking only one of the branches $(Z_\pm)$ we would be picking a preferred time orientation for our quantum state. Such a situation seems to be the relevant one for inflationary models, which explicitly break time-reversal symmetry.

\paragraph{Spontaneous CPT Breaking}
But how can we make sense of the sum of two partition functions? In general we cannot. However, in this case we can understand it as emerging from a \emph{phase transition}. Let us say we had access to a hypothetical UV completion of the $T^2$-deformed theory, whose partition function on $\Sigma$ we call $Z_\text{UV}$. In the deep UV, $Z_\text{UV}$ would be CPT-invariant. From the bulk we can understand this from the fact that in the limit of small volume of the slice $\Sigma$ we are probing the Euclidean part of the bulk. As we increase the volume of the slice, and correspondingly flow to the IR limit of $Z_\text{UV}$, a transition point is reached on the bulk side when enter the Lorentzian regime. At this point, the bulk picture can be either future or past-oriented, relative to the time-symmetric slice. In fact, because we started from a CPT-invariant picture, what happens is that we get a superposition of the two bulk time-orientations. Correspondingly, we get a splitting of $Z_\text{UV}$ into two partition functions $Z_\pm$, which are CPT-duals of each other. Secretly, their sum is still just $Z_\text{UV}$, just evaluated in the IR limit. We can think of this simply as a change of basis in theory space. What we have just described is an example of a field theory phase transition arising, in this case, from the \emph{spontaneous breaking of CPT symmetry}.\footnote{The author thanks Aron Wall for making this point so vividly clear to him over the years, in such an enthusiastic manner.} (See \cite{Araujo-Regado:2022gvw}, \cite{Wall:2021bxi} for similar points made in a different context) It is in this sense that the sum of $Z_\pm$ makes sense. The reader should contrast that with the point made in Section \ref{sectors} about the lack of a vector space structure on the space of theories. 

\paragraph{Contour-Superposition Correspondence}

An important conjecture of this paper is motivated in Section \ref{contour}. The choice of superposition in \eqref{branches} seems to be intimately related to the choice of class of geometries to sum over in the gravitational path integral. We explored this hypothesis in minisuperspace, where the class of geometries reduces to a choice of integration contour for the lapse. The chosen contour determines which saddle-points dominate the path integral giving a behaviour that matches a particular superposition in \eqref{branches}. It would be interesting to obtain an explicit map between a given linear combination and the corresponding contour, taking into account Piccard-Lefschetz theory \cite{Feldbrugge:2017kzv, DiazDorronsoro:2017hti} and the recent literature on the “allowability" of complex metrics \cite{Witten:2021nzp, Jonas:2022uqb}. For now, we just give the reader a taste of this connection by making a few identifications. After understanding fully how it works in the minisuperspace toy model, we believe this might give a new avenue to study the space of possible classes of geometries in the full gravitational path integral by studying properties of different linear combinations on the field theory side. The point is that, once a choice for the class of geometries is made, and hence a corresponding choice of field theory partition function $Z[g]$, we must stick to it.

\paragraph{A Taste of Branch Interference and Decoherence}
The aim of the holographic program is to recast all quantum gravitational quantities in terms of field theory objects. The holographic field theory is then conjectured to provide the \emph{definition} of quantum gravity. It is in this spirit that we should take the field theory partition function seriously and as the central object from which we can hope to learn about how quantum gravity actually works. Therefore, we believe that, given the partition function $Z[g]$, we should use that to compute expectation values in the theory. In field theory language these would be correlation functions. To take an example in minisuperspace, for concreteness, (where $Z[g]=Z(q)$ becomes a function of the scale factor $q$ only):
\bea
\expval{p_q...p_q}_q &=&\frac{1}{Z(q)}\frac{\partial^n}{\partial q^n}Z(q),\label{exp}\\
&=&\frac{1}{Z(q)}\frac{\partial^n}{\partial q^n}\left(A_+Z_+(q)+A_-Z_-(q)\right),\\
&=&\frac{A_+Z_+(q)}{Z(q)}\left(\frac{1}{Z_+(q)}\frac{\partial^n}{\partial q^n}Z_+(q)\right)+\frac{A_-Z_-(q)}{Z(q)}\left(\frac{1}{Z_-(q)}\frac{\partial^n}{\partial q^n}Z_-(q)\right),\\
&=&\mathcal{A}_+(q)\expval{p_q...p_q}_q^++\mathcal{A}_-(q)\expval{p_q...p_q}_q^-,
\eea
where $q=a^2$ is the scale factor on $S^2$ and $p_q$ is its conjugate momentum. Similar expressions can be written beyond minisuperspace. We have contributions from both branches with weighting factors $\mathcal{A}_+(q), \;\mathcal{A}_-(q)\in\mathbb{C}$. They have the property that $\mathcal{A}_+(q)+\mathcal{A}_-(q)=1$ but they cannot be interpreted as probabilities unless they are real and positive, which is not generically the case. 

We can make a bit more progress in the CPT-invariant subspace of quantum gravity theories, where the states are given by a subset of allowed branch superpositions, namely that of \eqref{CPTsubspace}. In those cases we have that:
\bea
\mathcal{A}_\pm(q)=\frac{1}{2}+\frac{1}{2}i \tan{\left[\arg{\left( Z_\pm(q)\right)}\pm \varphi\right]},
\eea
where $\varphi$ is a relative phase between the two branches. But the phase of each branch, $\arg{\left( Z_\pm(q)\right)}$, varies with $q$. Equation \eqref{solution} in fact suggests that it oscillates with a period of order:
\bea
\frac{q}{-\mu}\sim \frac{q}{-\mu}+O\left(\frac{1}{|c|}\right),\label{period}
\eea
where $\mu<0$ is the dimensionful deformation parameter and $c$ is the central charge of the $d=2$ CFT we had to deform. This becomes increasingly small in the semiclassical limit, which corresponds to $|c|\to\infty$. In this regime, if we average \eqref{exp} over some range of $q$ at least of order $|\mu|O\left(1/|c|\right)$ (noting that $\expval{p_q...p_q}_q^\pm$ varies slowly with $q$ by comparison with the phase\footnote{As can be checked through explicit computation from \eqref{solution}.}), we get:
\bea
\overline{\expval{p_q...p_q}}_q \approx 
\frac{1}{2}\expval{p_q...p_q}_q^++\frac{1}{2}\expval{p_q...p_q}_q^-,
\eea
which we can interpret as equal $50\%$ contributions from the future and past directed branches and we thus recover a good notion of probability. This equal split is what we intuitively expect if we demand CPT invariance. The emergence of probabilities hints at the quantum mechanical decoherence of the branches. In the presence of additional matter, averaging over a small amount of $q$ is physically and operationally meaningful since the scale factor is a proxy for time \cite{DeWitt:1967yk}, which works better and better in the regime $\frac{q}{-\mu}\gg \frac{1}{|c|}$. Every physical experiment is conducted in a non-vanishing time interval, one that is increasingly longer than the oscillation period \eqref{period} the closer we are to the semiclassical regime. It would be interesting to study whether and how a good probability measure on the space of possible branches emerges when we include matter. We believe the framework presented here and its future extensions might give some insight into what it means to do quantum mechanics of closed systems \cite{Hartle:1992as}.

\paragraph{Space of Solutions}

It is worth noting that the formalism of Cauchy Slice Holography uniquely specifies the RG flow trajectory but it does not fully specify the CFT that we need to start with. What it does is specify the dependence of $Z_\text{CFT}$ on the Weyl factor, through the anomaly equation. However, it does not fix the functional dependence within a conformal class of the spatial metric. Thus, many possible partition functions can seemingly be used as a starting point for solving the Hamiltonian constraint. This point was also subsequently made in \cite{Chakraborty:2023yed}. However, one would expect that further consistency conditions should be imposed on the CFT if appropriate semiclassical gravity results are to be recovered in the bulk, like correlation functions. 

Even more crucially, the CFT must at the very least include in its spectrum an operator of the form:
\bea
:\left(\Pi_{ab}\Pi^{ab}-\frac{1}{d-1}\Pi^2\right):,
\eea\label{protected}
which has scaling dimension \emph{exactly} $2d$! This is a highly non-trivial requirement since this dimension is not protected by conformal symmetry and in general picks up anomalous contributions. This condition will greatly reduce the allowed starting CFTs and it follows purely from imposing the Hamiltonian constraint away from the large volume limit. In $d=2$ this is just the $T\Bar{T}$ operator, which is well-defined (and has dimension $4$) for any QFT, at least in flat space. Although we are working on a curved space, we might guess that this requirement will be less restrictive in the $d=2$ case.

But in $d=2$ we have the further simplification that the only physical degree of freedom of the spatial metric is the Weyl factor, and so the anomaly equation fully specifies $Z_\text{CFT}$. The interpretation of this in the bulk is related to the fact that pure gravity in $2+1$-dimensions is topological, allowing for less freedom in specifying physically distinct bulk states.\footnote{This is no longer true if one includes matter. Then there is an infinite family of functional dependencies on the Weyl factor and correspondingly infinitely many more bulk states, modulo the caveat made in the previous paragraph.} This is the reason why, in $d=2$, Cauchy Slice Holography already gives us the full space of \emph{explicit} solutions to the Hamiltonian constraint.

\paragraph{The Need for UV Completion}
Equation \eqref{branches} defines the quantum gravity state over the whole superspace of metric configurations on $\Sigma$ via the deformed partition function defined in \eqref{deformation}. However, the fact that the deforming operator \eqref{def} is irrelevant means that this theory has a cutoff beyond which correlation functions cannot be trusted. In order to probe smaller scales we need to UV complete this theory. Because the deformation is guaranteed to be along an RG flow line, when we go to the IR limit of this theory, we end up back at the original CFT we started with. From the gravity perspective, probing smaller scales corresponds to taking the volume of $\Sigma$ to be smaller, which semiclassically is equivalent to going “back in time". So this gives us a field theoretic approach to the early Universe. Notice how the usual dS/CFT story \cite{Strominger:2001pn} is recovered in the IR limit of Cauchy Slice Holography.  In minisuperspace we had the advantage that the deforming operator was well-defined everywhere along the flow so we could already obtain some quantum “gravity" information.

\paragraph{Uniqueness of Cosmological State}
One of the most drastic conclusions from the framework of Cauchy Slice Holography is the fact that, given a holographic theory $Z[g]$, the quantum gravity theory thus defined has a \emph{unique} allowed state satisfying the constraints.\footnote{Ignoring for now the issue of the normalization factor raised by equation \eqref{normalization}.} This is a consequence of the slice $\Sigma$ on which we are defining the state having no boundary. Had it had a boundary we would have had freedom in specifying the boundary conditions to the holographic partition function. We have no such freedom on closed slices. This point was emphasized in Section \ref{contour}. 

A criticism to this idea could be that maybe there are many bulk states, but it is simply the case that each has a different holographic dual. See Section \ref{sectors} for an extended discussion on this point. 

The very deep and interesting question remains of how to do quantum mechanics when the observer is part of the quantum system. In particular, how are we to obtain a notion of Gibbons-Hawking entropy in the semiclassical limit for an observer constrained to the static patch of an emergent dS spacetime? This seems to require the emergence of a finite-dimensional “Hilbert space"\footnote{The use of scare quotes is to remind the reader that the work of \cite{Chandrasekaran:2022cip} suggests that one should instead think in terms of the algebra of observables restricted to the static patch. They argue that it is a von Neumann algebra of Type $\text{II}_1$ which means it does \emph{not} have a faithful irreducible representation in a Hilbert space. Yet a von Neumann entropy can be defined.} associated to a subregion from the one-dimensional Hilbert space associated to the full closed slice. This is not the usual pattern we see in entangled composite systems. We leave the difficult problem of obtaining this directly from the unique quantum state \eqref{branches} (with matter included) as an open question. However, we believe that our framework can give a new avenue to explore this problem by focusing attention on the field theory. 

It would be very interesting to understand how Cauchy Slice Holography is consistent with the edge mode story in gravity \cite{Donnelly:2016auv}. From the perspective of the field theory, if we make a cut on the $S^2$, there is a Hilbert space living there, which may somehow be related to the codimension-2 boundary degrees of freedom of the bulk theory. Could it be the case that all that there is for an observer in the Universe are edge states on their cosmological horizon?

\paragraph{Bulk and Boundary Unitarity}
In Section \ref{CPT} we analyzed the unitarity properties of both quantum gravity and the dual field theory. The latter turns out to lack reflection-positivity. Still, this holographic description allows for a very simple proof of bulk unitarity\footnote{In this paper we only discuss unitarity of the quantum theory on the full closed slice $\Sigma$. A very important question is whether unitarity is preserved once we restrict attention to the quantum theory of a spatial subregion. In \cite{Wall:2021bxi} it was argued, in the context of AdS/CFT, that restricting to subregions breaks bulk unitarity, even if the full quantum theory is unitary. We leave applying this analysis to closed cosmologies for future work. This will have major implications for the quantum description of gravity from the perspective of an observer in the static patch.}, which seems to hold for any choice of class of geometries $\mathcal{C}$ in the definition of the transition amplitude \eqref{dynamical}. Of course, in light of the uniqueness comments above, bulk unitarity is rendered essentially trivial. In fact, it was the pursuit of a positive-definite inner product which gave more evidence towards the uniqueness statements.

\paragraph{Role of CPT}
Motivated by the CPT-theorem in QFT, we studied the meaning of a CPT transformation in quantum gravity and how imposing bulk CPT-invariance restricts the class $\mathcal{C}$ and correspondingly the allowed superpositions of branches in \eqref{branches} to the set in \eqref{CPTsubspace}. According to the work of \cite{Susskind:2023rxm}, which argues that CPT should be viewed as a gauge symmetry in quantum gravity, we should restrict our attention to the CPT-invariant sector of the theory. Similarly, in the spirit of the discussion in an earlier paragraph, another motivation to restrict to CPT-invariant states could be that of making sense, from the field theory side, of the sum of the two CPT-dual field theory branches, $Z_\pm$, as emerging from a phase transition. Nevertheless, our analysis so far seems to suggest that there is no analogue of the CPT-theorem for quantum gravity, since the proof of unitarity does not rely on CPT symmetry. An interesting open question is how, once matter is included, the CPT-theorem for QFT in curved spacetime emerges from the underlying quantum gravity theory.

\paragraph{What about Baby Universes?}

In Section \ref{unitarity} we constructed an inner product on the space of ``metric eigenstates", given by \eqref{holography}. This followed from the fact that the gravitational path integral generates solutions to the Hamiltonian constraint, plus demanding Hermiticity of the inner product. Because the spatial slices are closed, the dual field theory description factorizes. This leads to an apparent tension with the fact that, from the bulk side, the gravitational path integral should sum over all geometries interpolating between the two boundaries, which include spacetimes that connect between the two ends. It would be interesting to study the relationship between the formalism of Cauchy Slice Holography hereby presented and the idea of ensemble averaging over the couplings of the dual field theory (see \cite{Saad:2021uzi}). Can this be made consistent with the requirement (discussed earlier) that we need the dimension of the quadratic operator in \eqref{protected} to be protected? We leave for future work the resolution to this important question.

\paragraph{Acknowledgements:}
The author is supported by a Harding Distinguished Postgraduate Scholarship and by the AFOSR grant FA9550-19-1-0260 “Tensor Networks and Holographic Spacetime”. The author is thankful to the following people for support and discussions: Aron Wall, Bilyana Tomova, Rifath Khan, Prahar Mitra, Santiago Agui-Salcedo, Philipp Hoehn, Manus Visser, Ronak Soni, Yiming Chen, Vasudev Shyam, Marija Tomasevic and Ayngaran Thavanesan. 

\appendix

\section{Minisuperspace Operators}

Here we derive the appropriate differential operators in minisuperspace.

For this we start with the Lorentzian Einstein-Hilbert action written in the ADM formalism:
\bea
I_\text{grav}=\frac{1}{16\pi G_N}\int dt\int_\Sigma N\sqrt{g}\left(K_{ab}K^{ab}-K^2+R-2\Lambda\right)=\text{KE}-\text{PE},
\eea
where this is true for the case of no spatial boundary, i.e. $\partial\Sigma=\emptyset$, otherwise get an extra Gibbons-Hawking term.
The extrinsic curvature is defined to be:
\bea
K_{ab}=\frac{1}{2}\mathcal{L}_n g_{ab}=\frac{1}{2N}\left(-\frac{\partial g_{ab}}{\partial t}+N_{(a|b)}\right),
\eea
where $n$ is the unit normal to the equal-time slices.

In minisuperspace ($N_a=0,\;\;g_{ab}=q\;\Omega_{ab})$, in 2+1 bulk dimensions, for $\Sigma=S^2$, we get:
\bea
K_{ab}=-\frac{1}{2N}\frac{\Dot{q}}{q}g_{ab}.
\eea
So the kinetic part of the gravitational action becomes:
\bea
I_\text{grav}\supset -\frac{1}{2}\frac{4\pi}{16\pi G_N}\int dt \frac{\Dot{q}^2}{Nq}.
\eea
We get the expected negative contribution to the kinetic energy coming from the conformal mode $q$.

The momentum conjugate to $q$ is defined to be:
\bea
p_q:=\frac{\delta I_\text{grav}}{\delta \Dot{q}}=-\frac{4\pi}{16\pi G_N}\frac{\Dot{q}}{Nq}=-i\frac{\partial}{\partial q}.
\eea

Now, before we go to minisupersapce, the conjugate momentum to the metric $g_{ab}$ on $\Sigma$ is defined to be:
\bea
\Pi^{ab}:=\frac{\delta I_\text{grav}}{\delta \Dot{g}_{ab}}=-\frac{\sqrt{g}}{16\pi G_N}(K^{ab}-Kg^{ab})=-i\frac{\delta}{\delta g_{ab}}.
\eea
We can now write this in terms of the minisuperspace momentum $p_q$:
\bea
\Pi^{ab}=\frac{1}{2}\frac{\sqrt{\Omega}}{4\pi} g^{ab}\;q \;p_q=-i\frac{1}{2}\frac{\sqrt{\Omega}}{4\pi} g^{ab}\;q \;\frac{\partial}{\partial q}.
\eea
The global scale transformation on $\Sigma$ is given by:
\bea
\int_{S^2}\Pi=-iq\frac{\partial}{\partial q},
\eea
as expected.

\iffalse
Now let us do it for the matter field. The Lorentzian action is:

\bea
I_\text{matter}=-\frac{1}{2}\int_\mathcal{M}\sqrt{\textbf{g}}\left(\textbf{g}^{\mu\nu}\partial_\mu\Phi\partial_\nu\Phi+m^2\Phi^2\right).
\eea
The conjugate momentum is:
\bea
\Pi_\Phi:=\frac{\delta I_\text{matter}}{\delta \Dot{\Phi}}=-\sqrt{\textbf{g}}\left(\textbf{g}^{00}\Dot{\Phi}+\textbf{g}^{0a}\partial_a\Phi\right)=-i\frac{\delta}{\delta \Phi}.
\eea

In minisuperspace ($\Phi(x,t)=\Phi(t)$), the kinetic part of the action is:
\bea
I_\text{mater}\supset \frac{1}{2}4\pi \int dt \;\frac{q\;\Dot{\Phi}^2}{N},
\eea
so we get a positive contribution to the kinetic energy from matter.
The momentum conjugate to the minisuperspace variable $\Phi$ is:
\bea
p_\Phi:=\frac{\delta I_\text{mater}}{\delta\Dot{\Phi}}=4\pi \frac{q\;\Dot{\Phi}}{N}=-i\frac{\partial}{\partial\Phi},
\eea
so we get:
\bea
\Pi_\Phi=\frac{\sqrt\Omega}{4\pi}p_\Phi=-i\frac{\sqrt\Omega}{4\pi}\frac{\partial}{\partial\Phi}.
\eea
We get that:
\bea
\int_{S^2}\phi\Pi_\phi=-i\phi\frac{\partial}{\partial\phi},
\eea
as expected.
\fi

%\subsection{Conventions}

\bibliographystyle{ieeetr}
\bibliography{references}

\end{document}